\documentclass[twocolumn]{aastex631}

\usepackage{hyperref}
\usepackage{subfigure}
\usepackage{graphicx}
\usepackage[flushleft]{threeparttable}
\usepackage{soul}

\begin{document}

\title{Evolution of the Inner Accretion Flow in Swift~J1727.8--1613 across Intermediate States: Insights from Broadband Spectral and Timing Analysis}

\correspondingauthor{Swadesh Chand}
\email{swadesh.chand@iucaa.in; swadesh.chand@mx.nthu.edu.tw}

\author[0000-0003-3499-9273]{Swadesh Chand}
\affiliation{Inter-University Centre for Astronomy and Astrophysics, Pune, Maharashtra-411007, India}
\affiliation{Institute of Astronomy, National Tsing Hua University, Hsinchu 300044, Taiwan}

\author[0000-0002-0333-2452]{Andrzej A. Zdziarski}
\affiliation{Nicolaus Copernicus Astronomical Center, Polish Academy of Sciences, Bartycka 18, PL-00-716 Warszawa, Poland}

\author[0000-0003-1589-2075]{Gulab C. Dewangan}
\affiliation{Inter-University Centre for Astronomy and Astrophysics, Pune, Maharashtra-411007, India}

\author[0009-0005-4510-4051]{Pragati Sahu}
\affiliation{Department of Pure and Applied Physics, Guru Ghasidas Vishwavidyalaya (A Central University), Bilaspur (C. G.)—495009, India}


\begin{abstract}

We present a comprehensive broadband spectral and variability study of the newly detected black hole X-ray binary Swift~J1727.8--1613 in the intermediate states during its 2023 outburst, using multi-mission observations from NICER, NuSTAR, AstroSat, and Insight-HXMT. Spectral data up to 78 keV in the hard-intermediate state (HIMS) require models with two Comptonizing regions. In contrast, models with a single Comptonizing region adequately describe the soft-intermediate states (SIMS), implying a significant evolution in the disk-corona geometry between the states. The hard X-ray tail above $100$ keV in the HIMS, detected with both AstroSat/CZTI and Insight-HXMT/HE, indicates that the electron population in the corona is not purely thermal but rather hybrid, with a power-law distribution above the thermal cutoff. While both the reflection modeling and disk continuum fitting favor a truncated disk geometry in the HIMS, the disk in the SIMS moves substantially closer to the innermost stable circular orbit, accompanied by a significant rise in disk temperature. This interpretation is further supported by the increase in the QPO frequency from $\sim1.3$ to $\sim6.6$ Hz. From joint modeling of the disk continuum and reflection component and assuming the distance of 3.4 kpc, we estimate a black hole mass of $10.3^{+5.5}_{-2.5}~M_\odot$, spin of $0.79^{+0.07}_{-0.15}$, and the disk inclination angle of $\sim37\degr$–-$53\degr$, which match well with the previously reported spectro-polarimetric measurements. We find a weakly variable or stable disk and a highly variable Comptonized component.

\end{abstract}


%% and \citet commands to identify citations. The citations are

\section{Introduction} \label{sec:intro}
Transient low-mass black hole X-ray binaries (BHXRBs) exhibit substantial evolution in their accretion geometry, particularly in the corona and in the size and shape of the accretion disk, across different spectral states during sporadic outbursts. These spectral state transitions are typically traced using the hardness–intensity diagram (HID), which follows a characteristic q-shaped track based on the changes in the source hardness with count rates in an anti-clockwise manner \citep{Homan2001ApJS..132..377H, Belloni2005A&A...440..207B, HomanBelloni2005Ap&SS.300..107H, Fender2009MNRAS.396.1370F, Belloni2010LNP...794...53B, Belloni2011BASI...39..409B}. During a typical outburst, BHXRBs spend most of their time in either the low/hard state (LHS) or the high/soft state (HSS), with relatively brief excursions through the short-lived HIMS and the SIMS.

The accretion geometry in the HSS is now well understood, with a standard optically thick and geometrically thin accretion disk \citep{Shakura1973A&A....24..337S} extending close to the innermost circular orbit (ISCO). This state is characterized by thermal disk emission peaking at $\approx1$ keV and low levels of X-ray variability. In contrast, the LHS is dominated by Comptonized hard X-ray emission from a hot corona and exhibits substantial variability. However, the geometry of the inner accretion flow in the LHS remains debated, particularly regarding whether the disk extends to the ISCO or is truncated at a large radius by a radiatively inefficient inner hot flow \citep[][for a review]{2006ApJ...653..525M, 2008ApJ...679L.113M, 2006MNRAS.367..659D, 2008MNRAS.387.1489R, 2010MNRAS.407.2287D, 2014MNRAS.437..316K, Bambi2021SSRv..217...65B}.

The HIMS and SIMS are transitional phases in which significant contributions from both disk and coronal emission are observed in the spectrum. The HIMS typically shows relatively strong Comptonized emission and strong variability, with a common detection of type-C quasi-periodic oscillations \citep[QPOs;] []{Casella2004A&A...426..587C, Casella2005ApJ...629..403C}. As the mass accretion rate increases, a source starts transitioning towards the SIMS, and the disk moves close to the ISCO \citep{Done2007A&ARv..15....1D, Belloni2011BASI...39..409B}. The observed spectrum is then dominated by thermal emission from the disk, and the source variability is weaker than in the HIMS. Type-B QPOs, which are relatively broader than Type-C QPOs, are sometimes detected in SIMS. Hence, the study of these intermediate states offers crucial insights into the dynamics of the inner accretion flow.  However, BHXRBs do not always evolve through all four canonical spectral states during an outburst. In some cases, the source remains in the LHS throughout the entire outburst, likely due to a rapid change or insufficient rise in the mass accretion rate. Such events are referred to as hard-only or failed outbursts \citep{Capitanio2009MNRAS.398.1194C, Stiele2016MNRAS.460.1946S, Sahu2024ApJ...975..165S}.

The hard X-ray emission observed in BHXRBs is commonly attributed to thermal Comptonization, arising from the inverse-Compton scattering of soft-disk photons by hot electrons in the corona. The electron distribution is predominantly thermal or Maxwellian at a mildly relativistic temperature and shows a sharp cutoff at $\approx100$ keV, particularly in the LHS \citep{Done2007A&ARv..15....1D, Zdziarski2020MNRAS.492.5234Z}. However, growing observational evidence for the presence of hard X-ray tails extending well beyond the thermal cutoff in several BHXRBs in the intermediate and soft spectral states suggests that the coronal electron population also includes a significant fraction of non-thermal electrons, with a power-law distribution \citep{McConnell2000ApJ...543..928M, McConnell2002ApJ...572..984M, Wardzinski2002MNRAS.337..829W, Cadolle2006A&A...446..591C, Poutanen2009ApJ...690L..97P, Jourdain2012ApJ...744...64J, Cangemi2021A&A...650A..93C, Cangemi2023A&A...669A..65C}. Similar non-thermal hard X-ray tail has also been reported in the LHS of the bright low-mass BHXRB MAXI~J1820+070 \citep{Zdziarski2021ApJ...914L...5Z}. Such a hybrid electron distribution, comprising thermal and non-thermal electron populations, can significantly modify the shape of the Comptonized spectrum \citep{Poutanen2009ApJ...690L..97P, Gierlinski1999MNRAS.309..496G, Zdziarski2001ApJ...554L..45Z}. While Comptonization from a thermally distributed electron population dominates the low-energy spectrum, the high-energy excess beyond the thermal cutoff can be described by Comptonization associated with the non-thermal electron distribution. Moreover, the non-thermal electron population can enhance electron-positron pair production, thereby introducing a thermostat effect that regulates and potentially lowers the coronal electron temperature \citep{Zdziarski2021ApJ...914L...5Z}.

Another critical component of the X-ray spectra of BHXRBs is the reflection emission, which arises due to the reflection of hard coronal X-rays by the disk. The reflection spectrum comprises mainly the iron $K_\alpha$ emission line, peaking at 6--7 keV, photoelectric absorption at low energies, and the reflection hump at 20--30 keV, originating from Compton scattering of high-energy photons. Gravitational redshift and the Doppler effect can distort the shape of the iron $K_\alpha$ line if the reflecting material is close to the ISCO \citep{Fabian1989MNRAS.238..729F}. Modeling this reflection spectrum is a primary approach for estimating the black hole spin, independent of mass and source distance. Another widely used method for estimating black hole spin is the disk continuum modeling, which requires prior knowledge of the black hole mass and source distance. Both methods rely on the assumption that the accretion disk is at the ISCO or very close to it, a configuration typically observed in the soft state. Furthermore, the presence of a high-energy non-thermal tail in the soft state irradiates the inner accretion disk. It produces relativistic reflection features, including a broadened Fe $K_\alpha$ line, whose degree of broadening depends on the black hole spin. Joint modeling of the relativistic reflection features and the thermal disk continuum, whose inner radius is linked to the ISCO, helps break degeneracies between different parameters, such as black hole mass, distance, and disk inclination angle, thereby improving constraints on the black hole spin \citep{Parker2016ApJ...821L...6P, Chand2022ApJ...933...69C, Zdziarski2024ApJ...962..101Z, Zdziarski2024ApJ...967L...9Z}. A detailed overview of the different methods of spin estimation is provided in a recent review paper by \citet{Zdziarski26}.

The bright low-mass X-ray transient Swift~J1727.8--1613 was first detected on 2023 August 24 \citep{Kennea2023GCN.34540....1K, Negoro2023GCN.34544....1N, Page2023GCN.34537....1P}, with the source flux reaching a peak of $\sim7$ Crab in the $15–50$ keV band \citep{Palmer2023ATel16215....1P}. Optical follow-up observations identified the source as a low-mass BHXRB with an orbital period of about $\sim10$ hours \citep{Mata2025A&A...693A.129M}. The distance was estimated to be $D=3.4\pm0.3$ kpc of \citet{Mata2025A&A...693A.129M} and $5.5^{+1.4}_{-1.1}$ kpc by \citet{Burridge25}. On the other hand, \citet{zdziarski2025ApJ...986L..35Z} argued for the distance $\lesssim$5 kpc by considering the Eddington limit on the bolometric hard-state luminosity. Here, we adopt the distance of \citet{Mata2025A&A...693A.129M}, but consider the scalings of our derived parameters with $D$. 

Based on reflection modeling in the LHS, \citet{Peng2024ApJ...960L..17P} reported a nearly maximally spinning black hole having a mass of $\sim10M_\odot$ and a disk inclination of $\sim40\degr$. Later, \citet{Svoboda2024ApJ...966L..35S} estimated the black hole spin to be $0.87\pm0.03$ by jointly modeling the disk continuum and reflection component in the HSS. Based on polarimetric measurements, the authors suggested an inclination angle of $30\degr$--$50\degr$. Additionally, substantial variation in the degree of polarization across spectral states, implying evolution in the inner accretion flow, is observed in this source \citep{Veledina2023ApJ...958L..16V, Svoboda2024ApJ...966L..35S}. The polarization studies also suggest that the corona is radially extended in the disk plane and lies perpendicular to the jet axis \citep{Veledina2023ApJ...958L..16V, Ingram2024ApJ...968...76I}. 

Then, \cite{Liu2024arXiv240603834L} showed that the broadband spectrum of Swift~J1727.8--1613 in the LHS is complex and requires Comptonization from a hybrid electron distribution. While the authors found that the contribution from non-thermal electrons was small in the LHS, it increased as the source approached the spectral transition. The QPO frequencies were also found to be evolving from $\sim 1.4$ to $\sim 2.6$ Hz over the HIMS \citep{Nandi2024MNRAS.531.1149N} and $\sim 0.2$ Hz to $\sim 1.86$ Hz over the transitioning period from the intermediate to the soft states \citep{Chatterjee2024ApJ...977..148C}. Similar evolution in the QPO frequencies from $\sim 0.3$ Hz in the LHS to $\sim 7$ Hz in the HIMS, along with a shrinkage in the coronal shape and disk truncation radius from 30--$40 ~\rm r_g$ to $10~ \rm r_g$, was also reported using NICER observations \citep{Rawat2025A&A...697A.229R}. While the QPOs in the HIMS were reported to be of type C \citep{Rawat2025A&A...697A.229R}, later investigations by \cite{Jin2025arXiv251010353J} suggested that they comprise a mixture of both C and B types.

In this work, we carry out a detailed broadband spectral and timing study of the newly discovered low-mass BHXRB Swift~J1727.8--1613 using multi-mission observations from NICER, NuSTAR, AstroSat, and Insight-HXMT during two distinct spectral states: the HIMS and SIMS. We demonstrate that the spectral data up to 78 keV in the HIMS requires models with two physically distinct thermal Comptonizing regions. Unlike previous studies in the LHS, our analysis of the extended broadband spectrum up to $200$ keV reveals that the hard X-ray tail in the HIMS becomes prominent only above $\sim100$ keV. This spectral behavior cannot be accounted for by thermal Comptonization alone and instead requires a hybrid Comptonization model that incorporates both thermal and non-thermal electron populations, underscoring the corona's complexity. We also show that the disk-corona geometry evolves significantly across the two observations. Using broadband X-ray data in SIMS, we estimate key parameters such as the black hole's mass and spin, and the disk inclination angle, using different accretion disk models and examine the model dependence of these parameters. In addition, we investigate the variability properties of the source in both HIMS and SIMS through detailed analyses of power density spectra (PDS) and rms–energy spectra. The paper is organized as follows. Section~\ref{sec:observation} describes the observations and data reduction. In Section~\ref{sec:Analysis and Results}, we present the results from our spectral and timing analysis. We discuss the implications of our findings in Section~\ref{sec:discussion}, and summarize our conclusions in Section~\ref{sec:conclusion}.

\section{Observations and Data Reduction} \label{sec:observation}

\begin{figure}
    \centering
    \includegraphics[width=\columnwidth]{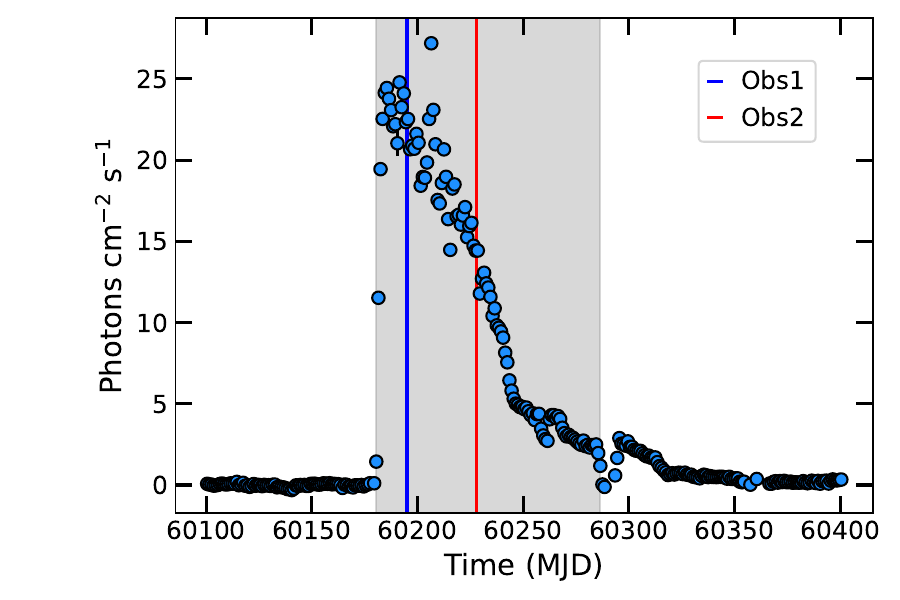}
    \caption{MAXI long-term lightcurve in the $2-20$ keV band during the outburst of Swift~J1727.8--1613 in 2023. The vertical lines indicate the positions of the two sets of observations considered in this work. The shaded region is the period that has been used to derive the HID in Figure~\ref{fig:hid_maxi}.}
    \label{fig:maxi_ltcrv}
\end{figure}

We analyze two distinct sets of observations of the recently discovered low-mass BHXRB SWIFT~J1727.8--1613, obtained during its 2023 outburst using NICER, NuSTAR, AstroSat, and Insight-HXMT. These datasets were obtained during the intermediate states (see below) and remain unexplored for broadband spectral modeling and detailed variability studies. The first dataset comprises simultaneous observations with NICER and NuSTAR. It is complemented by data from AstroSat and Insight-HXMT, providing broadband spectral coverage up to 200 keV near the peak of the outburst. The second dataset includes contemporaneous observations with NICER and NuSTAR, acquired during the decay phase of the outburst. Details of these observations are summarized in Table~\ref{tab:obs_ids}, and their positions within the outburst are indicated on the long-term MAXI light curve of the source, as shown in Figure~\ref{fig:maxi_ltcrv}. 
%%% MAXI long term lightcurve

%%% MAXI long term lightcurve
\begin{figure}
    \centering
    \includegraphics[width=\columnwidth]{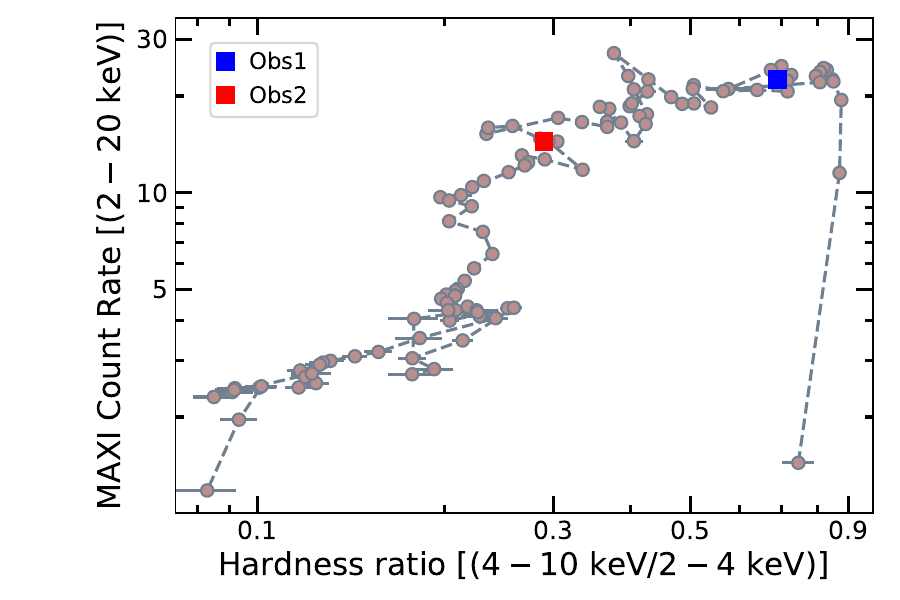}
    \caption{HID derived using MAXI observations (shaded region of Figure~\ref{fig:maxi_ltcrv}) of the 2023 outburst of Swift~J1727.8--1613. The squares mark the positions of the NICER observations used in this work, overplotted on the HID.}
    \label{fig:hid_maxi}
\end{figure}

%%%\subsection{HID}

To obtain an initial insight into the spectral states of the source during the two observations, we derived HID using the MAXI observations of Swift~J1727.8--1613 over the shaded interval in Figure~\ref{fig:maxi_ltcrv}, covering the entire 2023 outburst. We first calculated the hardness ratio (HR) by dividing the count rate in the $4$–$10$ keV band by that in the $2-4$ keV band. The HID was then obtained by plotting the HR against the MAXI count rate in the $2$–$20$ keV band. Figure~\ref{fig:hid_maxi} displays the resulting HID, which indicates that the source underwent a complete spectral transition during this outburst. The approximate positions of the two NICER observations analyzed in this study are overplotted on the HID based on their respective start times. These positions suggest that the source was likely in the HIMS during Obs1, characterized by an HR of $\sim0.7$, and possibly transitioned to the SIMS during Obs2, with the HR decreasing to $\sim0.3$.

\subsection{NICER}
We processed the NICER \citep{Gendreau2016SPIE.9905E..1HG} observations using the standard task \texttt{nicerl2} available within HeaSoft (v.6.34) and the most recent CALDB (v.xti\_20240206). To extract the time-averaged source and background energy spectra, we used the standard script \texttt{nicerl3-spec}, where the background spectra are estimated using the `3C50' model. We further used the \texttt{nicerl3-lc} task to generate the lightcurves from both the observations.

\subsection{NuSTAR}

The NuSTAR \citep{Harrison2013ApJ...770..103H} data were processed using the standard task \texttt{nupipeline}, which is part of the \texttt{NuSTARDAS} software package, along with the latest calibration files. Since the source was very bright during these observations, we applied the status expression \texttt{statusexpr=``(STATUS==b0000xxx00xxxx000)\&\& \hspace{1cm}(SHIELD==0)''} during data processing. Source and background spectra were then extracted considering circular regions of $70\arcsec$ radius, centered at the source position and an off-source region, respectively, using the standard task \texttt{nuproducts}. The corresponding response matrix files (RMFs) and auxiliary response files (ARFs) were generated using the same task. It is worth noting that the NuSTAR detectors may be affected by dead time{\footnote{It is the interval after each detected event during which the instrument processes the signal and cannot record new photons.}}, particularly for bright sources. While dead time does not have any effect on spectroscopy, it can significantly impact timing studies \citep[][for details on NuSTAR dead time]{Bachetti2015ApJ...800..109B, Bachetti2018ApJ...853L..21B}.

\subsection{AstroSat/CZTI}

AstroSat, India's first multi-wavelength astronomy satellite, was launched on September 28, 2015. It carries four co-aligned instruments onboard: the Soft X-ray Telescope \citep[SXT;][]{Singh2016SPIE.9905E..1ES, Singh2017JApA...38...29S}, the Large Area X-ray Proportional Counter Array \citep[LAXPC;][]{Yadav2016SPIE.9905E..1DY, yadav2016ApJ...833...27Y, Agrawal2017JApA...38...30A, Antia2017ApJS..231...10A}, Cadmium-Zinc-Telluride Imager (CZTI) \citep{Vadawale2016SPIE.9905E..1GV, Vibhute2021JApA...42...76V}, and the Ultra-Violet Imaging Telescope\citep[UVIT;][]{Tandon2017, Tandon2020}. For this study, we utilized data exclusively from the CZTI instrument. We downloaded the CZTI data from the AstroSat data archive{\footnote {\url{https://astrobrowse.issdc.gov.in/astro_archive/archive/Home.jsp}}} and produced clean event files using the latest version of the \texttt{cztpipeline} (v3.0){\footnote{\url{http://astrosat-ssc.iucaa.in/cztiData}}} and updated calibrations. This version of the \texttt{cztpipeline} includes an improved mask-weighting technique implemented in the \texttt{cztbindata} module \citep{Chattopadhyay2024ApJ...960L...2C}. The clean event files were then used to extract background-subtracted spectra for each quadrant using the standard tasks within \texttt{cztpipeline}{\footnote{\url{http://astrosat-ssc.iucaa.in/uploads/threadsPageNew_SXT.html}}} with the default inputs. For CZTI, background subtraction is inherent to spectral extraction and is based on analysis of long blank-sky observations and the mask-weighting technique. For typical exposures of less than $100$ ks, the background subtraction systematics are negligible. This becomes comparable to the statistical uncertainties only for exposures of $500$ ks and higher{\footnote {\url{http://astrosat-ssc.iucaa.in/uploads/meetings/13.\%20CZTISpecCalStatus_Mithun\%20-\%20Mithun\%20N\%20P\%20S.pdf}} (N. P. S. Mithun from the CZTI team, private communication)}. Finally, spectra from all four quadrants were combined to produce a high-signal-to-noise spectrum. We then added a 2\% systematic error to the CZTI merged spectral data to account for the instrument's calibration uncertainties. We analyzed the CZTI spectral data in the 40--200 keV band, excluding data below 40 keV because of the strong Xe hump, which peaks at $\sim30$ keV.

%%% Observation details
\begin{table*}
	\caption{List of Observations of SWIFT~J1727.8--1613}
 %  { log of observations}
   \centering
   \begin{tabular}{lcccccc}
      \hline
      \hline
    % & \multicolumn{2}{c}{$0.7-3$ keV} & \multicolumn{2}{c}{$3-10$ keV} 
    % \\\hline
    % & Epoch - 1 & Epoch - 2 & Epoch - 1 & Epoch - 2 \\
    Obs No & Instrument & Obs ID & Start Time & Stop Time & Eff. Exp.(s) \\
    &  & & (yyyy-mm-dd) & (yyyy-mm-dd) & \\
     \hline
        Obs1 & NICER & $6750010502$ & 2023--09--08 00:47:12 & 2023--09-08 13:13:35 & 2270  \\
           & NuSTAR/FPMA & $80902333008$ & 2023--09--08 00:56:09 & 2023--09--08 04:01:09 & 543.2  \\
           & FPMB &      \nodata         & 2023--09--08 00:56:09 & 2023--09--08 04:01:09 & 589.5 \\
           & AstroSat/CZTI & $9000005836$ & 2023--09--08 01:39:59 & 2023--09--09 00:23:28 & 26770\\
           & Insight-HXMT/HE & P061433800901 & 2023--09--08 02:13:25 & 2023--09--08 05:37:26 & 2235\\ \\
           
    Obs2 & NICER & 6203980136 & 2023--10--09 03:45:39 & 2023-10-09 16:16:20 & 1909 \\
       & NuSTAR/FPMA & 80902313016 & 2023--10--10 19:36:09 & 2023--10--10 23:06:09 & 1065 \\
       & FPMB & \nodata & 2023--10--10 19:36:09 & 2023--10--10 23:06:09 & 1138 \\
     \hline
   \end{tabular}
   \label{tab:obs_ids}
\end{table*}     

\subsection{Insight-HXMT/HE}

The first X-ray satellite of China, Insight-HXMT \citep{Zhang2014SPIE.9144E..21Z, Zhang2018SPIE10699E..1UZ, Zhang2020SCPMA..6349502Z}, launched on June 15, 2017, carries three different payloads onboard. They are: Low-Energy X-ray Telescope \citep[LE;][]{Chen2020SCPMA..6349505C}, Medium-Energy X-ray Telescope \citep[ME;][]{Cao2020SCPMA..6349504C}, and High-Energy X-ray Telescope \citep[HE;][]{Liu2020SCPMA..6349503L}. Here, we have used only HE data, which operates in the 20--250 keV band, for spectral analysis. We used HXMTDAS \citep[v.2.06;][]{Zhao2020ASPC..527..469Z} package to process the HE data following the standard procedure and default inputs. For generating source and background spectra and response files, the standard task \texttt{hpipeline} was used. The background is generated and scaled by the standard task \texttt{hebkgmap} within \texttt{hpipeline}, which employs the built-in background model by \citet{Liao2020JHEAp..27...14L} based on blank-sky observations. This background model introduces an average systematic error of 1.5\% in the background energy spectrum (26–-100 keV) for a typical exposure of 8 ks. The background systematic error varies in the range of 2\%--10\% above 120 keV \citep{Li2020JHEAp..27...64L}. For this study, we utilized the HE spectral data in the 30--150 keV band and discarded the data below $30$ keV due to calibration issues, and beyond 150 keV for significant background contamination. 

\section{Analysis and Results} \label{sec:Analysis and Results}
\subsection{Broadband Spectral Analysis}

\begin{figure}
    \centering
    \includegraphics[width=\columnwidth]{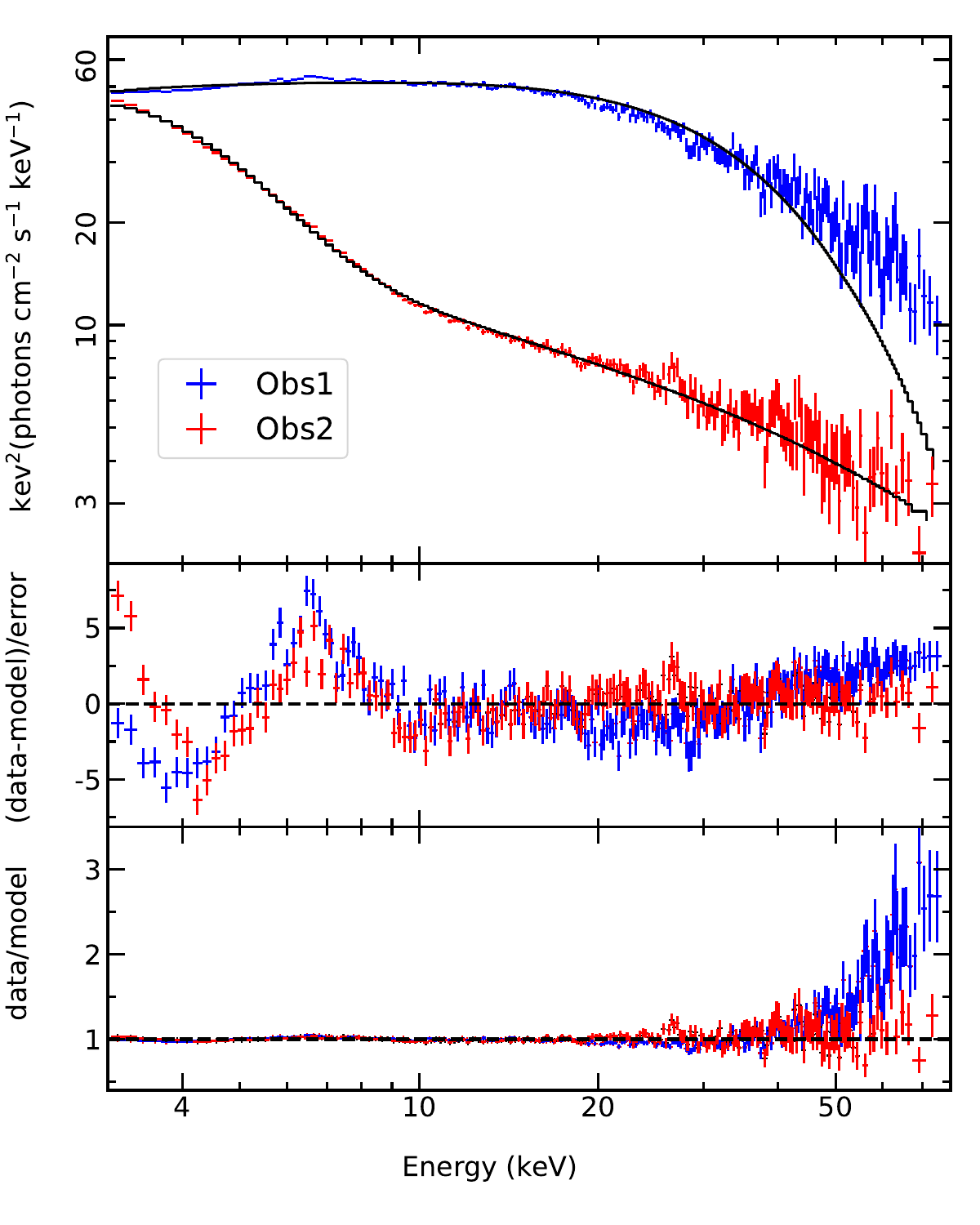}
    \caption{Upper panel: NuSTAR/FPMA spectral data in the HIMS (blue), the SIMS (red), and the fitted model \texttt{tbabs*(thcomp*diskbb)} (black solid line). Middle panel: data residuals with respect to the fitted model. Lower panel: data-to-model ratio. A prominent Fe $K_\alpha$ line and a weak reflection hump are observed.}
    \label{fig:residual_plots}
\end{figure}

%%% Xspec spectral plots for hard state data using M1
We performed broadband spectral analysis using XSPEC \citep[v.12.14.1;][]{Arnaud1996ASPC..101...17A} and quoted all the uncertainties on the best-fit parameters at the $90\%$ confidence ($\Delta \chi^2\approx2.71$) level. We applied the optimal binning algorithm \citep{Kaastra2016A&A...587A.151K}, with a minimum of 25 counts per bin, to the NICER, NuSTAR, and HE spectral data to enable the use of the $\chi^2$ statistic. For the CZTI spectral data, we retained the default binning criteria embedded in the \texttt{cztpipeline} \citep{Chattopadhyay2024ApJ...960L...2C, Chand2024ApJ...972...20C}. To account for any differences in the relative normalizations and possible calibration uncertainties between the different instruments, we multiplied a \texttt{plabs} component to our models. This component is purely phenomenological and does not alter the continuum's intrinsic shape or the key physical parameters. The parameters $\Delta \Gamma$ and $K$ of the \texttt{plabs} component were fixed at $0$ and $1$, respectively, for NuSTAR FPMA, while they were varied freely for the other instruments. We started the spectral analysis by jointly fitting the simultaneous (for Obs1) or contemporaneous (for Obs2) NICER and NuSTAR spectral data in the $0.7-78$ keV band with a model consisting of a multicolor disk blackbody component \citep[\texttt{diskbb};][]{Mitsuda1984PASJ...36..741M} to account for the thermal emission from the disk, and \texttt{thcomp} \citep{Zdziarski2020MNRAS.492.5234Z}, describing the spectra from Comptonization of the disk photons by thermal electrons in the corona. As \texttt{thcomp} is a convolution model, we extended the sample energy range from 0.01 to 1000 keV using the \texttt{energies} command in XSPEC. Furthermore, we treated the absorption by the interstellar medium using the Galactic absorption component \texttt{tbabs} with the abundances and cross sections taken from \cite{Wilms2000ApJ...542..914W} and \cite{Verner1996ApJ...465..487V}, respectively. This model \texttt{plabs*tbabs(thcomp*diskbb)} provided an inferior spectral fit and left prominent residuals at the Fe $K_\alpha$ band and reflection hump region, typically arising due to the reflection of the coronal irradiation from the accretion disk. Figure~\ref{fig:residual_plots} shows the NuSTAR/FPMA spectral data for the two different spectral states along with the data residuals to the model \texttt{tbabs(thcomp*diskbb)}. A clear difference in spectral properties between the two states is evident in this figure. The nature of the reflection hump also shows a transition between the two observations, with it appearing relatively prominent at high energies in Obs1 and relatively weak in Obs2.

%%% Residual Plots using NuSTAR/FPMA data

\subsubsection{HIMS}

%%% Reflection spectroscopy with Nicer and NuSTAR observation using Model-1 (double Comptonizing regions) for both the observations
\begin{table*}%[!ht]
	\caption{Best-fit broadband X-ray spectral parameters derived for Obs1 (HIMS)}
%	\tablenum{1}
   \centering
   \begin{tabular}{lccccc}
 
      \hline
      \hline
	Component & Parameter & M1 & M2 & M3 \\
        %& & & compps=thermal & compps=hybrid \\
        & & NICER/NuSTAR & NICER/NuSTAR & NICER/NuSTAR/HE/CZTI \\
	\hline
	TBabs & $N_{\rm{H}}$(10$^{22}$ cm$^{-2})$ & $0.30\pm0.02$ & $0.27\pm0.01$ & $0.27\pm0.01$ \\ \\

		Diskbb & $kT_{\rm{in}}$ (keV) & $0.39\pm0.02$ & $0.40\pm0.01$ & $0.39\pm0.01$ \\
		& norm ($\times10^5$) & $1.96^{+0.25}_{-0.75}$ & $1.54^{+0.49}_{-0.40}$ & $1.61^{+0.27}_{-0.29}$ \\ \\

        Thcomp & $\Gamma$ & $1.95^{+0.03}_{-0.09}$ & $1.92^{+0.04}_{-0.17}$ & $1.90\pm0.04$ \\
                & $kT_{\rm{e}}$ (keV) & $4.43^{+0.19}_{-0.21}$ & $4.60^{+0.26}_{-0.25}$ & $4.77\pm0.18$ \\
                & $C_{\rm{f}}$ & $0.45^{+0.03}_{-0.05}$ & $0.43^{+0.10}_{-0.13}$ & $0.40\pm0.04$ \\ \\
            	
	RelxillCp & $\Gamma$ & $1.80^{+0.05}_{-0.16}$  & \nodata & \nodata\\
                & $kT{\rm{{e}}}$ (keV) & $15.95^{+1.68}_{-0.93}$  & \nodata & \nodata\\
		    & $R_{\rm{in}}$ ($r_{\mathrm{g}}$) & $143.7^{p}_{-110.4}$  & \nodata & \nodata\\
		    & $\log \xi$ & $3.69^{+0.12}_{-0.11}$  & \nodata & \nodata\\
		    & $A_{\mathrm{Fe}}$ (solar) & $0.61^{+0.44}_{p}$  & \nodata & \nodata\\
                & $\log N$ & $15.98^{+0.51}_{p}$  & \nodata & \nodata\\
		      & $R_{\mathrm{ref}}$ & $0.64^{+0.44}_{-0.20}$  & \nodata & \nodata\\
		    & norm & $0.17^{+0.05}_{-0.03}$  & \nodata & \nodata\\ \\

        mbknpo  & $B$ & $1.0^{+0.12}_{p}$ & \nodata & \nodata \\ 
                & $I$ & $0.3^f$ & \nodata & \nodata \\ \\

        Relconv & $R_{\rm{in}}$ ($r_{\mathrm{g}}$) & \nodata & $249.5^{p}_{-196.4}$ & $212.6^{p}_{-168.8}$ \\ \\
                            
        Xilconv & $R_{\mathrm{ref}}$ & \nodata & $0.17^{+0.02}_{-0.03}$ & $0.15^{+0.04}_{-0.03}$ \\
                & $A_{\mathrm{Fe}}$ (solar) & \nodata & $1^f$ & $1^f$ \\
                 &$\log \xi$ & \nodata & $3.89^{+0.11}_{-0.09}$ & $3.84\pm0.08$ \\ \\

        Compps [thermal/hybrid] & $kT_{\rm{e}}$ (keV) & \nodata & $15.32^{+0.84}_{-0.58}$ & $13.73^{+0.32}_{-0.43}$ \\
                         & $\tau$ & \nodata & $9.7^{p}_{-2.2}$ & $9.96^{+1.19}_{-0.83}$ \\
                         & $\gamma_{\rm{min}}$ & \nodata & \nodata & $1.33\pm0.01$ \\
                        & norm ($\times10^4$) & \nodata & $5.7^{+2.3}_{-2.4}$ & $7.5^{+0.9}_{-1.5}$ \\ \\
                        
                         Flux$^{\rm{unabs}}_{\rm{0.001-500}~keV}$ [$10^{-7}{\rm{erg~cm^{-2}~s^{-1}}}$] & & $\sim3.3$ & $\sim3.3$ & $\sim3.4$ \\ \\ 
                           
                         \hline
        Cross-calibration & $\Delta \Gamma _{\rm{NICER}}$ & $0.08\pm0.01$  & $0.08\pm0.01$ & $0.08\pm0.01$ \\
                          & $K_{\rm{NICER}}$ & $0.86\pm0.02$  & $0.86\pm0.02$ & $0.86\pm0.02$ \\
                          & $\Delta \Gamma _{\rm{NuSTAR~B}}$ & $0.004^{+0.005}_{p}$  & $0.004^{+0.005}_{p}$ & $0.004^{+0.005}_{p}$ \\
                           & $K_{\rm{NuSTAR~B}}$ & $0.97\pm0.01$ & $0.97\pm0.01$ & $0.97\pm0.01$ \\
                           & $\Delta \Gamma _{\rm{HE}}$ & \nodata & \nodata & $<0.03$ \\
                           & $K_{\rm{HE}}$ & \nodata & \nodata & $0.94^{+0.01}_{-0.10}$ \\
                           & $\Delta \Gamma _{\rm{CZTI}}$ & \nodata & \nodata & $0.10^{+0.02}_{-0.04}$ \\
                           & $K_{\rm{CZTI}}$ & \nodata & \nodata & $0.63^{+0.06}_{-0.08}$ \\
                           \hline
		    & $\chi^2$/dof & $457.3/589$ & $458.7/592$ & $601.5/681$ \\
	\hline 
    \end{tabular}
    \label{tab:Obs1}
    \begin{tablenotes}
    \item Notes -- M1: \texttt{plabs*tbabs(thcomp*diskbb+mbknpo*relxillCp)}, M2: \texttt{plabs*tbabs*relconv*xilconv(thcomp*diskbb+compps[thermal])}, M3: \texttt{plabs*tbabs*relconv*xilconv*(thcomp*diskbb+compps[hybrid])}, where the spectral data from HE and CZTI were also added. See \cite{Svoboda2024ApJ...966L..35S} for the definition of the \texttt{mbknpo} in \texttt{xspec}. The spin of the black hole and disk inclination angle are fixed at $0.998$ and $40\degr$, respectively. $^f$ -- indicates fixed parameters. $p$ -- indicates that the parameter is pegged at the model-defined highest or lowest bound during error estimation.
    \end{tablenotes}
\end{table*}

 For Obs1, we then attempted to model reflection from ionized material by convolving \texttt{ireflect} \citep{1995MNRAS.273..837M} with \texttt{thcomp} and incorporating a \texttt{gaussian} component to account for the Fe $K_\alpha$ emission line. The \texttt{ireflect} model is the generalized convolution form of \texttt{pexriv}, and describes only the hard X-ray shape of the reflected spectrum, excluding all the line emissions. However, this model configuration, \texttt{plabs*tbabs(ireflect*thcomp*diskbb+gaussian)}, yielded a statistically poor fit, with $\chi^2/dof = 1634.4/596$, and left significant residuals at high energies. This indicates that a model involving a single Comptonizing region is insufficient to reproduce the observed broadband spectral shape, and an additional Comptonization component may be required to account for the high-energy excess. A requirement for two distinct Comptonizing regions to describe the broadband ($0.7-100$ keV) spectrum in the LHS of the low-mass BHXRBs GX~339–4 \citep{Chand2024ApJ...972...20C} and AT2019wey \citep{Sahu2025arXiv251106393S} has also been reported by our group. A similar behavior is observed in the LHS of MAXI~J1820+070, indicating that multiple Comptonization regions are required in the hard states of BHXRBs \citep{2021ApJ...909L...9Z, Zdziarski2021ApJ...914L...5Z}.
 
\begin{figure*}
    \centering
    \includegraphics[width=0.32\linewidth]{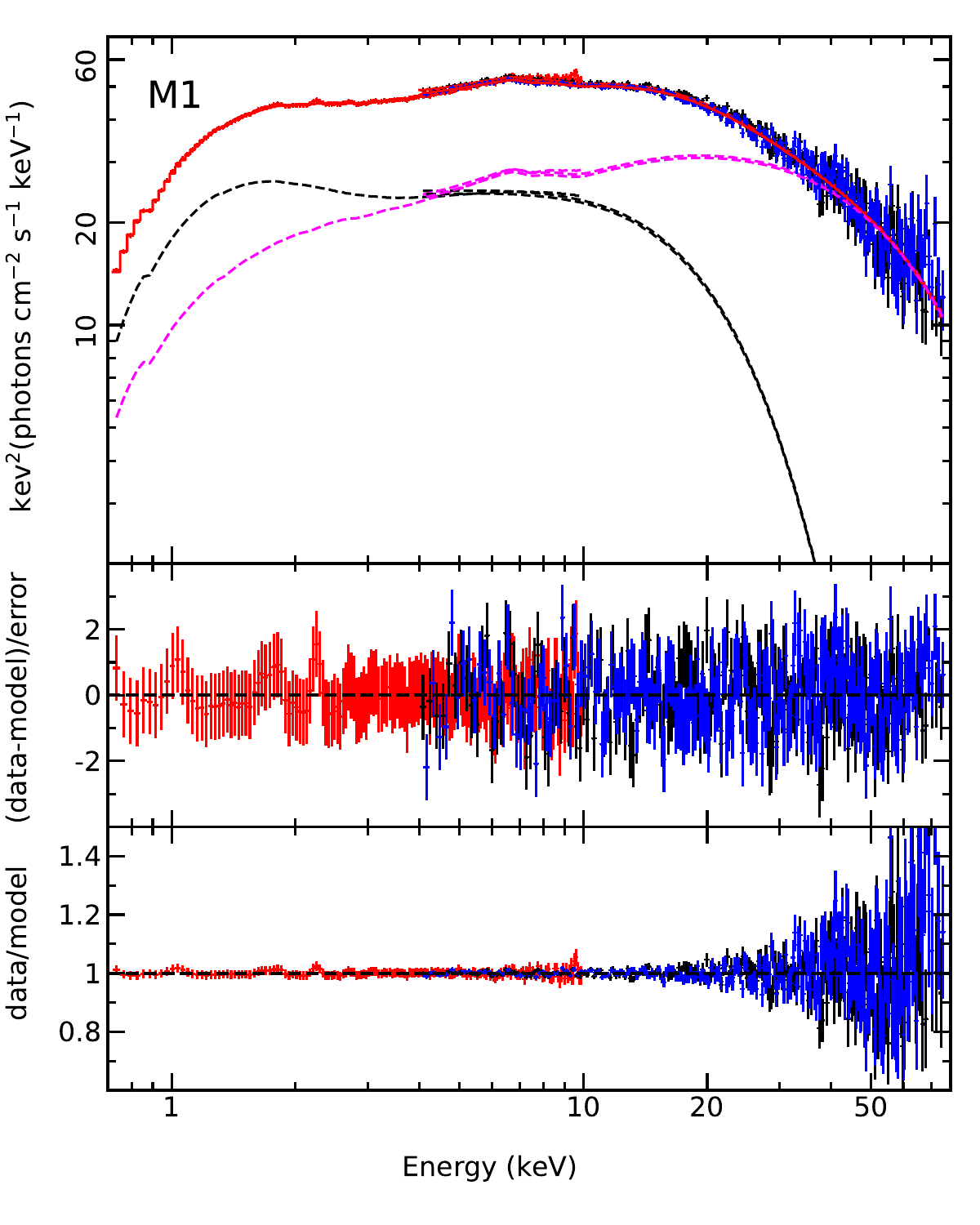}
    \includegraphics[width=0.32\linewidth]{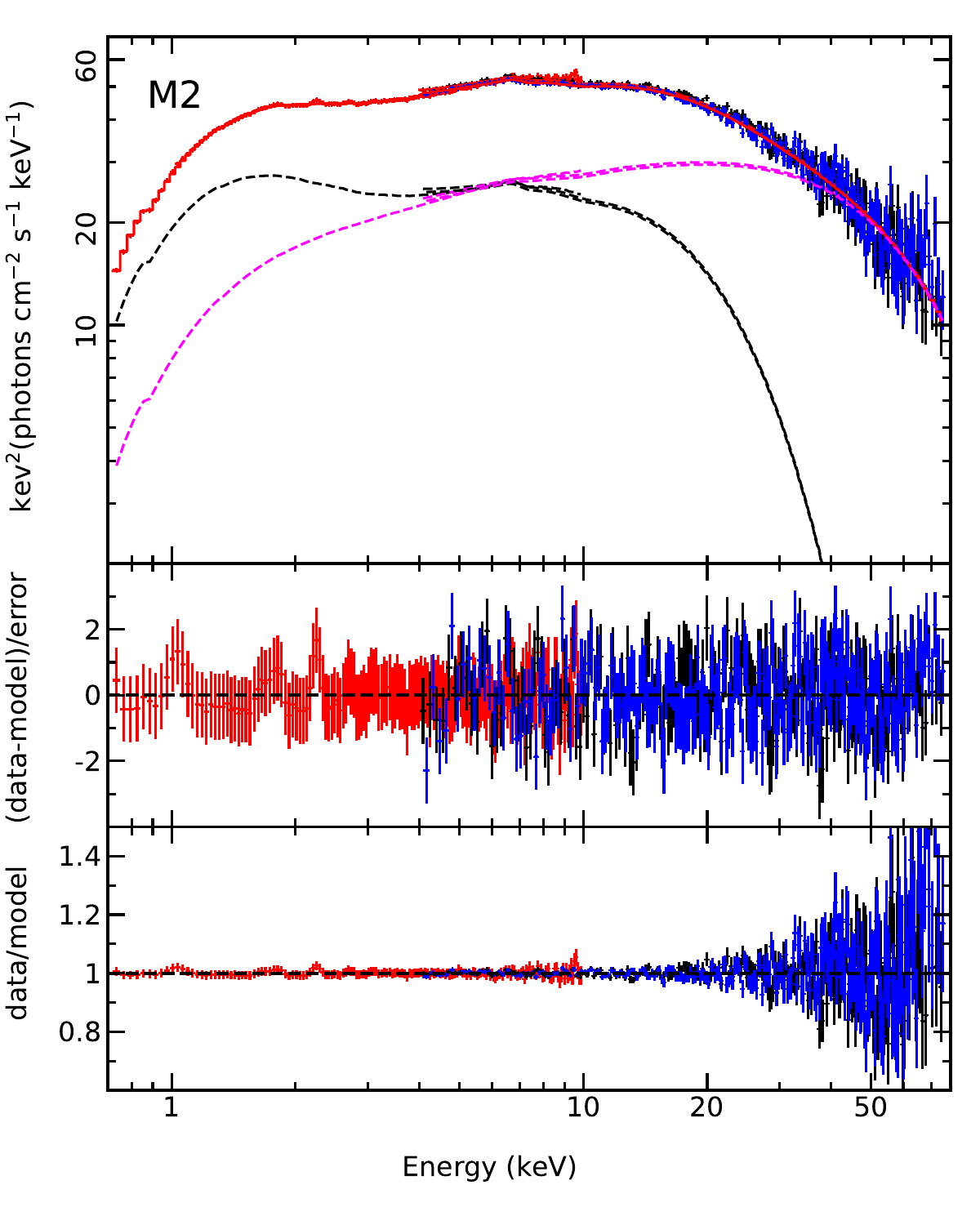}
    \includegraphics[width=0.32\linewidth]{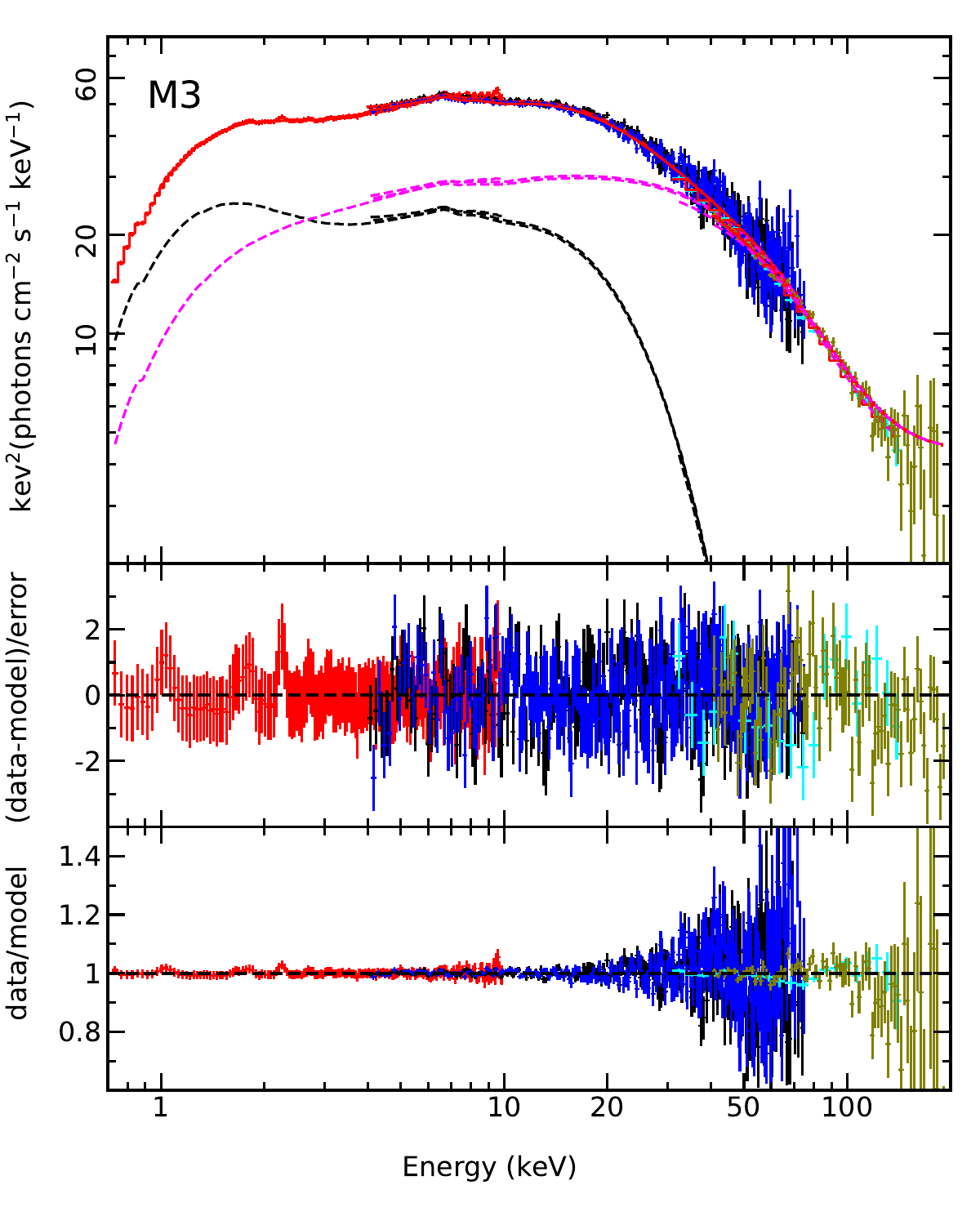}
    \caption{Joint NICER (red), NUSTAR/FPMA (black), and FPMB (blue) spectral data in the HIMS along with the best-fit models M1: \texttt{plabs*tbabs(thcomp*diskbb+mbknpo*relxillCp)} (left panel), and M2: \texttt{plabs*tbabs*relconv*xilconv(thcomp*diskbb+compps[thermal])} (middle panel). The right panel shows the joint NICER, NuSTAR/FPMA, FPMB, HE (cyan), and CZTI (olive) spectral data along with the best-fit model M3: \texttt{plabs*tbabs*relconv*xilconv*(thcomp*diskbb+compps[hybrid]).} The black dotted line indicates the \texttt{thcomp} convolved with \texttt{diskbb} component, whereas the magenta dotted line represents the \texttt{relxillCp} for M1, and \texttt{relconv} and \texttt{xilconv} components convolved with \texttt{compps} components for M2 and M3.}
    \label{fig:hardstate_Obs1}
\end{figure*}

 We then removed \texttt{ireflect} and \texttt{gaussian} components and adopted a more physically motivated model \texttt{relxillCp}, which is a part of the family of relativistic blurred reflection models \texttt{relxill} \citep{Dauser2014MNRAS.444L.100D, Garcia2014ApJ...782...76G}. The \texttt{relxillCp} model self-consistently computes the reflection spectrum arising from the irradiation of thermally Comptonized coronal emission from the disk and also includes relativistic effects due to strong gravity near the black hole. The shape of the incident radiation in \texttt{relxillCp} is the same as in the \texttt{nthcomp} model. In \texttt{relxillCp}, we fixed the black hole spin at the maximum value of $0.998$ and disk inclination angle at $40\degr$  \citep{Peng2024ApJ...960L..17P, Svoboda2024ApJ...966L..35S}. The outer disk radius was fixed at $1000~r_{\rm{g}}$, and a single emissivity profile ($\propto r^{-q}$, $q$ being the emissivity index) was considered throughout the whole disk with $q$ fixed at the canonical value of 3. The photon index ($\Gamma$) and electron temperature ($kT_{\rm{e}}$) were varied independently of those in \texttt{thcomp}. Other free parameters of \texttt{relxillCp} were the inner disk radius ($R_{\rm{in}}$), log of disk ionization ($\log\xi$), iron abundance ($A_{\rm{Fe}}$), log of disk density ($\log N$), reflection fraction ($R_{\rm{ref}}$), and normalization. To avoid the over prediction of the contribution of the reflection at lower energies, we multiplied \texttt{relxillCp} with a \texttt{mbknpo} component, which helps to put a break on the reflection spectrum below a certain energy, and thus preventing the unphysical runaway of the photon spectrum at low energies \citep{Steiner2024ApJ...969L..30S, Svoboda2024ApJ...966L..35S}. The \texttt{mbknpo} component is parameterized with the break energy $B$ and index $I$, used for the correction for energies $E<B$. We set these parameters according to the prescription outlined in \citet{Svoboda2024ApJ...966L..35S}. This composite model, \texttt{plabs*tbabs(thcomp*diskbb+mbknpo*relxillCp)} (hereafter M1), resulted in a significantly improved fit with $\chi^2/dof = 457.3/590$. The best-fit spectral parameters are listed in Table~\ref{tab:Obs1}, and the corresponding spectral models and data are shown in Figure~\ref{fig:hardstate_Obs1}. 

 The results from our model M1 clearly indicate that the broadband spectrum in the $0.7-78$ keV band in the HIMS requires two separate Comptonizing regions. The \texttt{relxillCp} component represents the hard Comptonizing region with $\Gamma \sim1.8$ and $kT_{\rm{e}} \sim16$ keV, likely having a spherically extended geometry. This region is primarily responsible for the origin of the reflection features at a distant part of the disk. On the other hand, the second Comptonizing region, described by \texttt{thcomp}, is found to have low electron temperature, $kT_{\rm{e}} \sim4$ keV, and exhibits a relatively steeper spectrum with $\Gamma \sim1.95$. This soft Comptonizing region may appear as a warm layer over the inner parts of the accretion disk. We also explored the possibility of an additional contribution to the reflection from the soft Comptonizing region by incorporating a second \texttt{relxillCp} component as a reflection-only component. However, we did not find any significant improvement in our spectral fitting. This implies that the reflection from the soft Comptonizing region is negligible and that the hard Comptonizing region dominates the overall reflection spectrum.

 The above model could not properly constrain the inner disk radius, $R_{\rm{in}}$. However, the lower bound at the $90\%$ confidence level indicates that the disk is truncated away from the ISCO by at least $\sim33~r_{\rm{g}}$ (where $r_{\rm{g}}=GM/c^2$ is the Gravitational radius). We also calculated the inner disk radius from the best-fit value of the \texttt{diskbb} normalization, following the XSPEC definition and applying the color correction factor of 1.7 \citep{Shimura1995ApJ...445..780S}. The estimated radius is $\sim29~r_{\rm{g}}$, which agrees closely with the value obtained from the reflection modeling. Furthermore, the temperature of the inner accretion disk is estimated to be $\sim0.4$ keV, typically higher than that observed in the LHS. The disk appears highly ionized, with slightly elevated density. Within the uncertainties, the iron abundance is consistent with the solar value. Additionally, the reflection fraction, $R_{\rm{ref}}$, defined as the ratio of the coronal emission reaching the disk and the observer, is found to be $\sim0.64$, which in turn also favors the disk truncation scenario. All the best-fit spectral parameters indicate that the source was in HIMS during Obs1.

We then modeled the hard Comptonization spectra with the \texttt{compps} \citep{Poutanen1996ApJ...470..249P} model by replacing \texttt{relxillCp} in M1. The Comptonization model \texttt{compps} can treat the electron distribution in a corona as purely thermal, purely non-thermal, or hybrid for different shapes of the coronal geometry. Here, we assumed the electron distribution in the corona as purely thermal and the coronal geometry as spherical (\texttt{geom=0}) in the \texttt{compps} model. We also tested other corona geometries but observed no significant change in the spectral fit. Therefore, we kept the geometry spherical, which also yields a faster method for fitting the spectra than the other geometries. In the \texttt{compps}, the seed photons for the Comptonization are considered to be multi-color disk photons provided by the \texttt{diskbb} component. We fixed the disk inclination angle at $40\degr$ and iron abundance at solar. The other free parameters of this model component are the electron temperature, $kT_{\rm{e}}$, and the optical depth, $\tau$, of the scattering medium. Furthermore, to account for disk reflection, we convolved the \texttt{xilconv} model with both Comptonizing regions. The reflection fraction, $R_{\rm{refl}}$, and log of disk ionization, $\log\xi$, parameters in \texttt{compps} were allowed to vary freely. We kept the iron abundance fixed at solar, consistent with the value obtained from model M1 within the uncertainty. We also incorporated the \texttt{relconv} model to treat the relativistic blurring effects, while keeping the relevant parameters the same as those used for \texttt{relxillCp} in model M1. This model, \texttt{plabs*tbabs*relconv*xilconv*(thcomp*diskbb+compps)} (referred to as M2), provides a statistically acceptable fit to the broadband spectrum, with  $\chi^2/dof = 458.7/592$. The best-fit spectral parameters are given in Table~\ref{tab:Obs1}, and the best-fit spectra along with model components are shown in Figure~\ref{fig:hardstate_Obs1}.

We find that the optical depth ($\tau$) of the hard Comptonizing region is $\sim9.7$. The accretion disk appears to be highly ionized, and the relativistic reflection fraction ($R_{\rm{ref}}$) is measured to be $\sim0.2$, which is slightly lower than the value obtained with model M1. This reduction may result from the inclusion of reflection contributions from both Comptonizing regions in the current model, which could dilute the apparent strength of the relativistic reflection component relative to that in model M1. Similar to our model M1, the lower bound on the inner disk radius at $90\%$ confidence level suggests that the disk is truncated away from the ISCO by at least $43 ~r_{\rm{g}}$. Besides, the other best-fit spectral parameters from M2 are in good agreement with those derived from M1.

%%% Residual Plots for HE and CZTI spectral data using only the thermal Comptonization model
\begin{figure}
    \centering
    \includegraphics[width=\columnwidth]{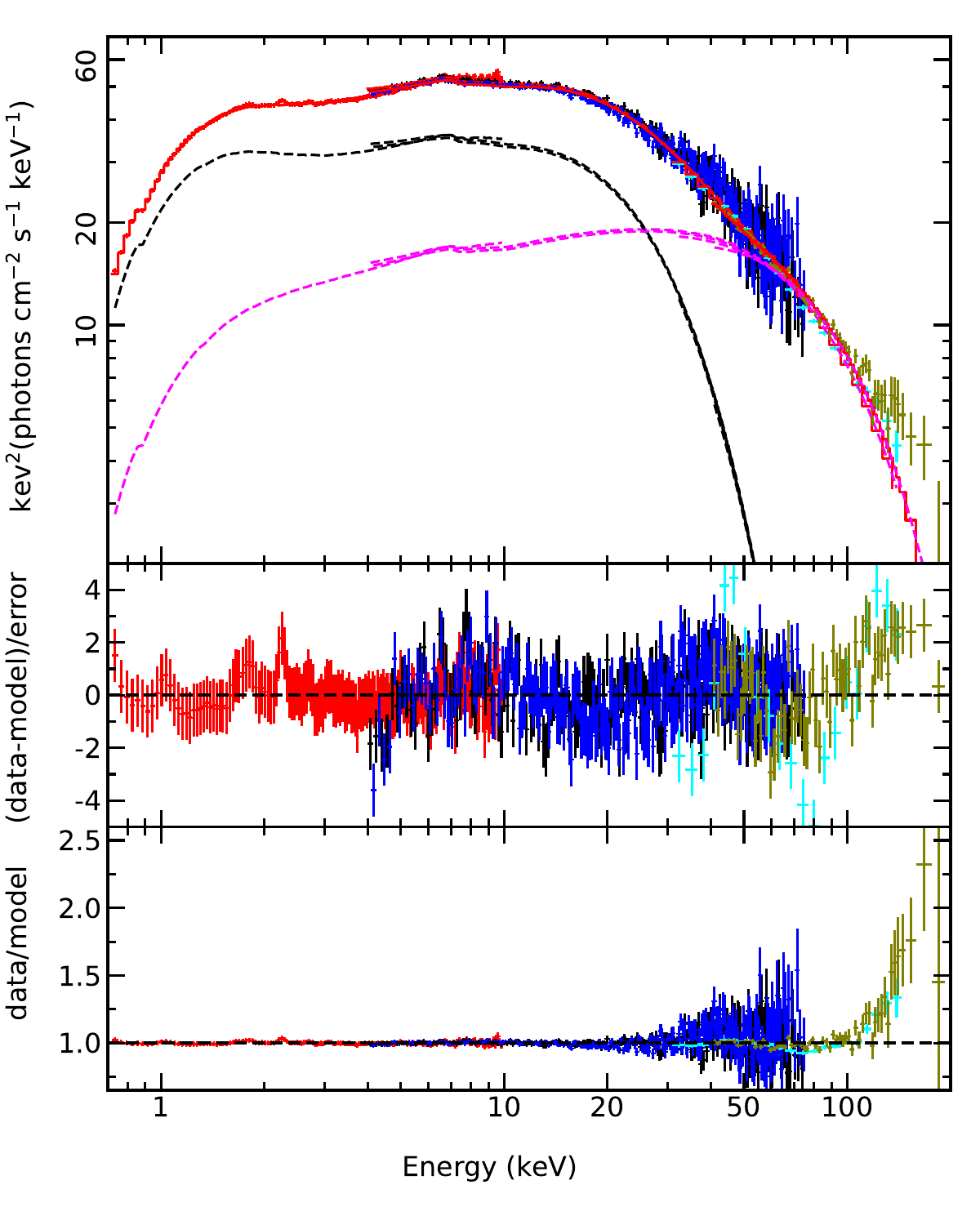}
    \caption{Upper panel: NICER (red), NUSTAR/FPMA (black), FPMB (blue), HE (cyan), and CZTI (olive) spectral data in the HIMS along with the fitted model M2, which considers only a Maxwellian distribution of the electrons in the corona. Middle and lower panels: deviations of jointly fitted NICER (red), NuSTAR/FPMA (black) and FPMB (blue), HE (cyan), and CZTI (olive) spectral data from the model M2. The black and magenta dotted lines have the same meaning as in Figure~\ref{fig:hardstate_Obs1} for model M2. A prominent excess above $100$ keV is observed. The data is rebinned for plotting purposes only.}
    \label{fig:res_he_czti}
\end{figure}

In the next scenario, we incorporated spectral data from the HE and CZTI instruments onboard Insight-HXMT and AstroSat, respectively, to achieve broadband spectral coverage up to 200 keV. We first assessed calibration uncertainties in these datasets by fitting them up to 78 keV, along with NICER and NuSTAR spectra, using both models M1 and M2. The resulting best-fit spectral parameters were either identical or consistent with those obtained from joint spectral fitting of NICER and NuSTAR data in the 0.7--78 keV band using models M1 and M2, indicating no significant cross-calibration discrepancies among the instruments. Furthermore, inspection of the HE spectral data confirms that the background is significantly lower than the source signal in the $100-150$ keV band, 
with source and background count rates of $96.7\pm0.2 $c/s and $48.1\pm0.9$ c/s, respectively. This suggests a detection significance of $\sim52 \sigma$ for the source emission in the $100-150$ keV band. We then extended the energy ranges of the HE and CZTI spectral data to $150$ keV and $200$ keV, respectively. However, we find that both models, M1 and M2, were unable to describe the broadband spectra up to $200$ keV and yielded poor spectral fits with $\chi^2$/dof$=803/679$ for M1 and $886.8/682$ for M2, leaving a prominent hump at high energies ($\gtrsim100$ keV). Figure~\ref{fig:res_he_czti} presents the broadband spectral data in the $0.7-200$ keV band fitted with the model M2, along with the residuals, highlighting the deviations between the observed data and the model.

The apparent excess at $\gtrsim100$ keV further suggests that models with thermal Comptonization and disk reflection alone are insufficient to account for the high-energy emission and may be attributable to a power-law tail in the distribution of hot coronal electrons. The electron distribution is \texttt{compps} hybrid, with the Lorentz factor ($\gamma_{\rm{min}}$) defining the transition point from a thermal to a power-law electron distribution. We allowed $\gamma_{\rm{min}}$ to vary freely while fixing $\gamma_{\rm{max}} = 1000$ and keeping the electron power-law index fixed at 2. This model, referred to as M3, which assumes a hybrid electron distribution with Maxwellian and non-thermal above a specific value of $\gamma_{\rm{min}}$, provides a perfect and significantly improved spectral fit in the $0.7-200$ keV band with $\chi^2$/dof$=601.5/681$ with $\Delta \chi^2 \approx -285$ for 1 d.o.f. The best-fit spectral parameters obtained from M3 are listed in Table~\ref{tab:Obs1}, and the best-fit spectral data with the model components are shown in Figure~\ref{fig:hardstate_Obs1}. We find $\gamma_{\rm{min}}$ to be $\sim1.33$, implying a significant contribution from the non-thermal distribution. The other best-fit spectral parameters obtained with model M3 are in close agreement with those from our earlier models, assuming a purely thermal electron distribution up to $78$ keV.

\subsubsection{SIMS}

For Obs2, we performed a joint spectral analysis using contemporaneous NICER and NuSTAR data in the 0.8--78 keV band. The NICER data below $0.8$ keV were discarded to avoid calibration uncertainties. These observations were taken during the decay phase of the 2023 outburst. Residuals from fitting the NuSTAR/FPMA spectral data with the model \texttt{tbabs(thcomp*diskbb)} revealed a moderately broad Fe K$\alpha$ line and a weak reflection hump (see Figure~\ref{fig:residual_plots}). We initiated the joint fitting by including a \texttt{plabs} component to account for calibration uncertainties, a \texttt{gaussian} component to model the iron line, and the \texttt{ireflect} component convolved with \texttt{thcomp} to account for reflection from ionized material. Our current model, \texttt{plabs*tbabs(ireflect*thcomp*diskbb+gaussian)}, provided a poor spectral fit with $\chi^2$/dof$=826.2/523$. We also observed that replacing \texttt{thcomp} with \texttt{cutoffpl} yielded a slightly improved fit, although the overall fit quality remained unsatisfactory. Despite poor spectral fits, a relatively higher inner-disk temperature, $kT_{\rm{in}}$, and a steeper photon index, $\Gamma$, suggest that the source may have gradually transitioned to the SIMS between Obs1 and Obs2.

%%% Reflection spectroscopy with Nicer and NuSTAR observation for soft state observation
\begin{table*}%[!ht]
	\caption{Best-fit broadband X-ray spectral parameters derived for Obs2 (SIMS)}
%	\tablenum{1}
   \centering
   \begin{tabular}{l@{\hskip 0.2in}c@{\hskip 0.2in}c@{\hskip 0.2in}cc }
 
      \hline
      \hline
	Component & Parameter & M4 & M5 & M6\\
	\hline
	TBabs & $N_{\rm{H}}$(10$^{22}$ cm$^{-2})$ & $0.26\pm0.01$ & $0.28\pm0.01$  & $0.31\pm0.01$ \\ \\

	Diskbb & $kT_{\rm{in}}$ (keV)   & $0.87\pm0.01$ & \nodata  & \nodata  \\
		& norm ($\times10^3$) & $9.5^{+1.4}_{-0.8}$ & \nodata  & \nodata \\ \\

        Thcomp & $\Gamma$  & $2.36\pm0.03$ & $2.33^{+0.10}_{-0.06}$ & $2.35^{+0.03}_{-0.04}$ \\
               & $kT_{\rm{e}}$ (keV)  & $998.2^{p}_{-500.4}$ & $668.5^{p}_{-554.6}$  & $995^{p}_{-502}$\\
               & $C_{\rm{f}}$ & $0.25^{+0.04}_{-0.06}$ & $0.2^{+0.2}_{-0.1}$ & $0.23^{+0.05}_{-0.06}$ \\ \\

        Kerrbb & $a$ & \nodata & $0.79^{+0.07}_{-0.15}$  & \nodata \\
             & $i\degr$   & \nodata & $46.6^{+6.7}_{-9.2}$  & \nodata \\
             & $\dot{M}_{\rm d}$ ($\times10^{18}~\rm{g/s}$) & \nodata & $1.5\pm0.2$  & \nodata \\ 
             & $M_{\rm{BH}}$ ($M_\odot$) & \nodata & $10.3^{+5.5}_{-2.5}$  & \nodata \\
             & $D$ (kpc) & \nodata & $3.4^f$ & \nodata \\ \\

        BHSPEC [$\alpha=0.1$]  & $M_{\rm{BH}}$ ($M_\odot$) & \nodata & \nodata & $10.2^{+1.3}_{-0.4}$ \\
                & $L/L_{\rm{Edd}}$ & \nodata & \nodata & $0.12^{+0.02}_{-0.01}$ \\
                & $a$                   & \nodata & \nodata & $0.80^{p}_{-0.03}$ \\ 
                & $i\degr$              & \nodata & \nodata & $47.2\pm2.3$ \\
                & $D$ (kpc)               & \nodata & \nodata & $3.4^f$ \\ \\
             
        RelxillCp (refl. only) & $a$ & $0.98^{p}_{-0.29}$ & \nodata  & \nodata \\
                & $i\degr$  & $40^f$ & \nodata & \nodata \\
                & $R_{\rm{in}}$ ($r_{\mathrm{isco}}$)  & $1.7^{+0.9}_{p}$ & $1^f$ & $1^f$ \\
                & $q$ & $3.7^{+0.6}_{-0.4}$ & $6.6^{p}_{-3.0}$ & $6.0^{+2.1}_{-1.2}$ \\
		    & $\log \xi$  & $4.09^{+0.04}_{-0.11}$ & $4.26^{+0.12}_{-0.21}$ & $4.20^{+0.04}_{-0.10}$ \\
		    & $A_{\mathrm{Fe}}$ (solar)  & $1^f$ & $1^f$ & $1^f$ \\
                & $\log N$  & $17^f$ & $17.0\pm0.4$ & $17^f$ \\
		    & norm  & $0.19^{+0.01}_{-0.02}$ & $0.19\pm0.05$ & $0.21^{+0.02}_{-0.03}$ \\ \\

        mbknpo & $B$ & $1.57\pm0.07$ & $1.66\pm0.08$ & $1.54^{+0.07}_{-0.09}$ \\
               & $I$ & $0.8^f$ & $0.8^f$ & $0.8^f$ \\ \\

            %Flux$_{\rm{total}}$ [$10^{-7}{\rm{erg~cm^{-2}~s^{-1}}}$] & & $1.8$ & $1.8$\\ \\
            Flux$^{\rm{unabs}}_{\rm{0.001-500}~keV}$ [$10^{-7}{\rm{erg~cm^{-2}~s^{-1}}}$] & & $\sim2.5$ & $\sim2.5$ & $\sim2.6$ \\ \\

      \hline
        Cross-calibration & $\Delta \Gamma _{\rm{NICER}}$   & $0.13\pm0.01$ & $0.14\pm0.01$ & $0.13\pm0.01$\\
                          & $K_{\rm{NICER}}$   & $0.82\pm0.02$ & $0.81^{+0.01}_{-0.02}$  & $0.81\pm0.02$ \\
                          & $\Delta \Gamma _{\rm{NuSTAR~B}}$  & $0.002^{+0.006}_{p}$ & $0.003^{+0.006}_{p}$ & $0.002^{+0.006}_{p}$ \\
                           & $K_{\rm{NuSTAR~B}}$ & $0.98\pm0.01$ & $0.98\pm0.01$ & $0.98\pm0.01$ \\
                          \hline
		    & $\chi^2$/dof & $506.5/521$ & $470.8/520$ & $489/521$ \\
	\hline 
    \end{tabular}
    \label{tab:Obs2_softstate}
    \begin{tablenotes}
    \item Note. M4: \texttt{plabs*tbabs(thcomp*diskbb+mbknpo*relxillCp)}, M5: \texttt{plabs*tbabs(thcomp*kerrbb+mbknpo*relxillCp)}, M6: \texttt{plabs*tbabs(thcomp*bhspec+mbknpo*relxillCp)}. $^f$ -- indicates fixed parameters. $p$ -- indicates that the parameter is pegged at the model-defined highest or lowest bound during error estimation.
    \end{tablenotes}
\end{table*}

\begin{figure*}
    \centering
    \includegraphics[width=0.32\linewidth]{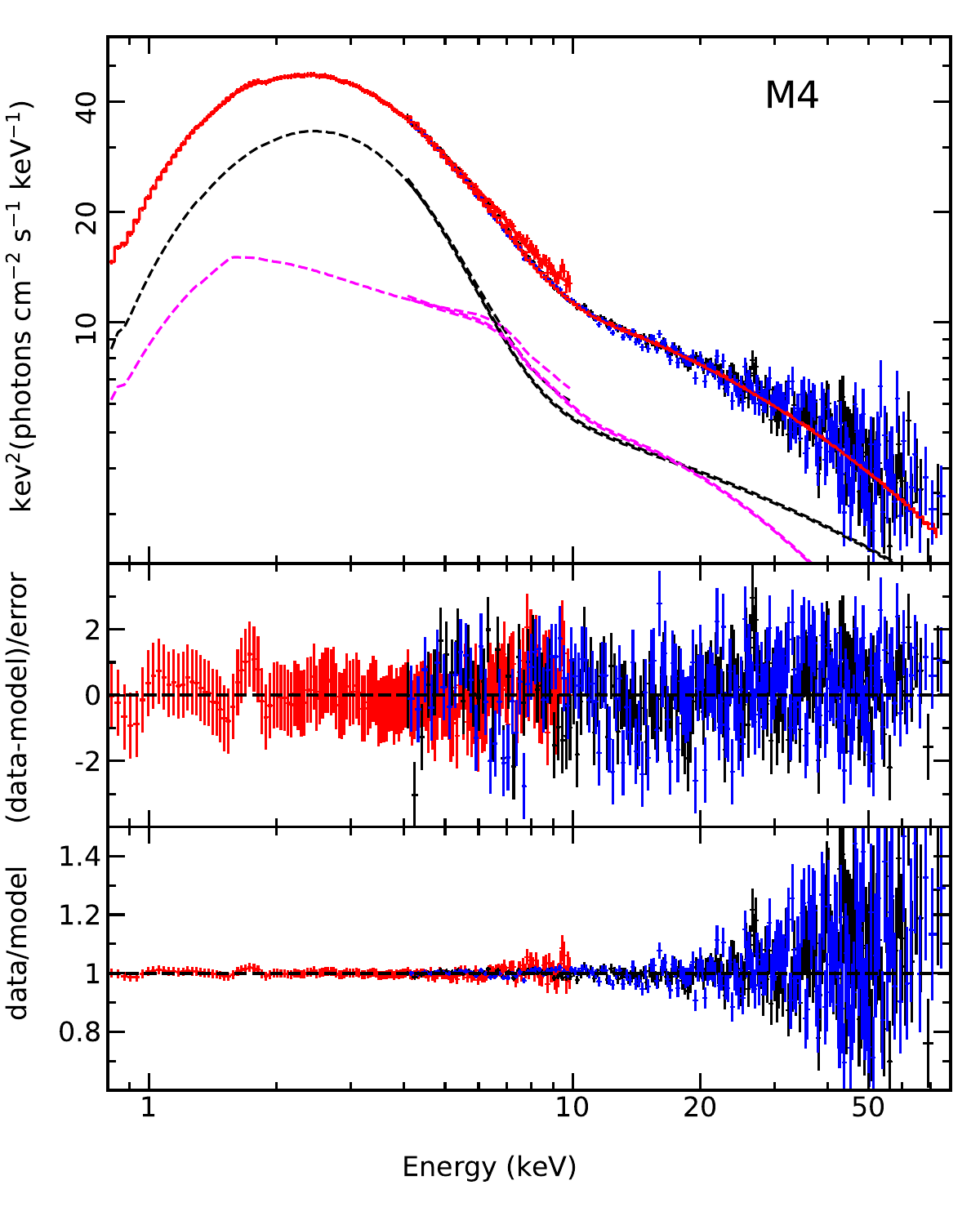}
    \includegraphics[width=0.32\linewidth]{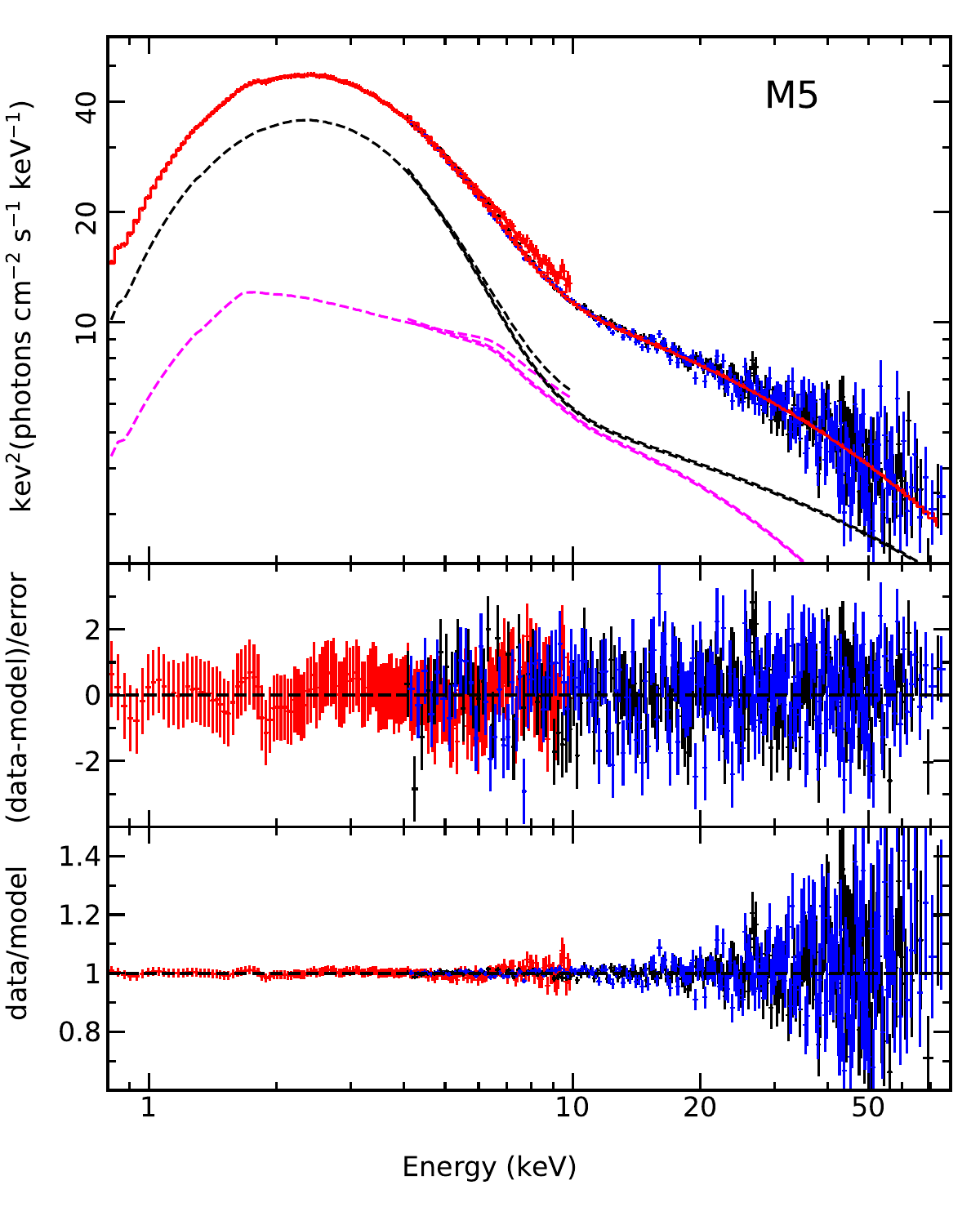}
    \includegraphics[width=0.32\linewidth]{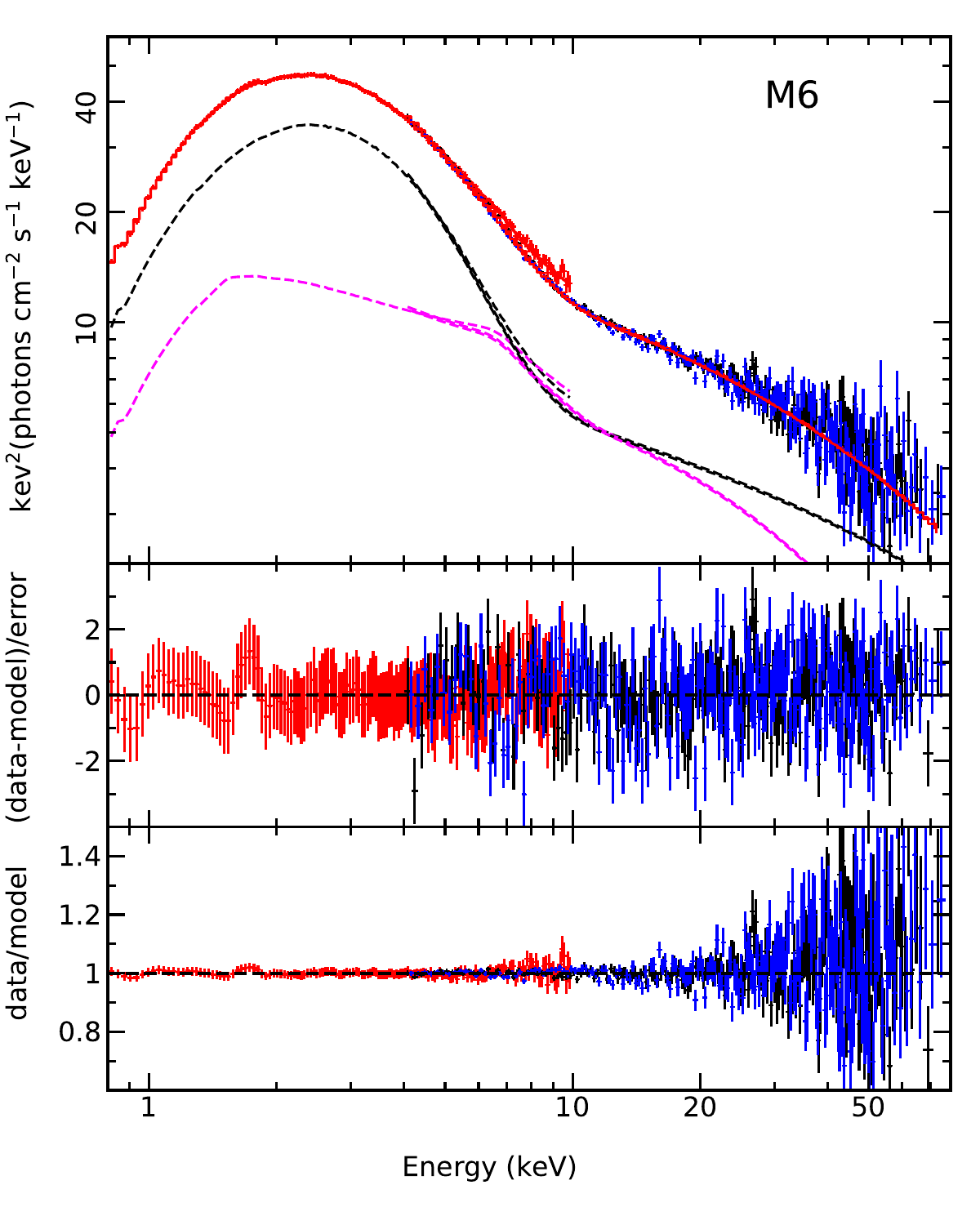}
    \caption{Joint NICER (red), NUSTAR/FPMA (black), and FPMB (blue) spectral data of the SIMS observations along with best-fit model M4: \texttt{plabs*tbabs(thcomp*diskbb+mbknpo*relxillCp)}, M5: \texttt{plabs*tbabs(thcomp*kerrbb+mbknpo*relxillCp)}, and M6: \texttt{plabs*tbabs(thcomp*bhspec+mbknpo*relxillCp)}. The black dotted line represents the \texttt{thcomp} component convolved with \texttt{diskbb} for M4, \texttt{kerrbb} for M5, and \texttt{bhspec} for M6, whereas the magenta dotted line shows the \texttt{relxillCp} (reflection only) component.}
    \label{fig:softstate_Obs2}
\end{figure*}

We then employed the more physically motivated and relativistically blurred reflection model, \texttt{relxillCp}, to perform broadband reflection spectroscopy, replacing the \texttt{ireflect} and \texttt{gaussian} components. Here, \texttt{relxillCp} is considered only as a reflection component by fixing the reflection fraction parameter at $-1$. Common parameters, such as $\Gamma$ and $kT_{\rm{e}}$, were tied between the \texttt{relxillCp} and \texttt{thcomp} components. The inclination angle and the iron abundance were also kept fixed at $40\degr$  and solar, respectively. Other free spectral parameters of the \texttt{relxillCp} model are the emissivity index ($q$), log of disk ionization ($\log\xi$), log of disk density ($\log\xi$), and normalization. Additionally, we varied the black hole spin ($a$) and the inner disk radius ($R_{\rm{in}}$) independently, accounting for the SIMS. The resulting model \texttt{plabs*tbabs(thcomp*diskbb+relxillCp)} (now referred to as M4) provides a good and statistically acceptable fit with $\chi^2/dof=506.5/521$. This further suggests that a model consisting of a single Comptonizing region, its associated reflection, and disk emission is sufficient to describe the $0.8$–$78$ keV spectrum in the SIMS, unlike the HIMS, where two distinct Comptonizing regions were required. The best-fit spectral parameters obtained from M4 are listed in Table~\ref{tab:Obs2_softstate}, and the corresponding spectral data and models are shown in Figure~\ref{fig:softstate_Obs2}.

The Galactic absorption column density, $N_{\rm{H}}$, is found to be $\sim0.26\times10^{22}\rm{cm^{-2}}$. The best-fit values of $kT_{\rm{in}}$ and $\Gamma$ indicate that the inner accretion disk has become significantly hotter ($\sim0.9$ keV), and the continuum has steepened, consistent with the source being in the SIMS. The electron temperature, $kT_{\rm{e}}$, could not be constrained, likely due to the absence of a clear high-energy cutoff. Compared to the HIMS, the inner disk radius has decreased substantially, reaching close to the ISCO with $R_{\rm{in}} \lesssim 2.6~R_{\rm{isco}}$. Additionally, the estimated radius of $\sim7-8~r_{\rm{g}}$ from the \texttt{diskbb} normalization following the similar procedure adopted in the HIMS also suggests a relatively closer extent of the disk. The disk ionization parameter is also found to be higher than that in the HIMS. Because disk density could not be constrained, it was fixed at its best-fit value of $10^{17}~\rm{cm^{-3}}$, which is slightly higher than the value inferred during HIMS. These spectral parameters together suggest that the source transitioned to the SIMS during Obs2.

We next replaced the \texttt{diskbb} component with the relativistic disk model \texttt{kerrbb} \citep{Li2005ApJS..157..335L}, which describes a geometrically thin, steady-state accretion disk around a Kerr black hole. The inner disk radius was fixed at the ISCO, and a zero-torque boundary condition was assumed at the inner edge. The spectral hardening factor was set to the canonical value of 1.7 \citep{Shimura1995ApJ...445..780S}, and disk self-irradiation was included. The black hole spin, $a$, and disk inclination angle, $i$, were tied between the \texttt{kerrbb} and \texttt{relxillCp} components and allowed to vary. Other free parameters of \texttt{kerrbb} are the disk mass accretion rate ($\dot{M}_{\rm d}$), and mass of the black hole ($M_{\rm{BH}}$). 

We note that the \texttt{kerrbb} model is based on the thin-disk approximation; therefore, the fitted spectrum is insensitive to the Eddington ratio (which, in reality, would affect the disk scale height). For a given spin, this approximation yields spectra that depend only on the disk's maximum temperature, $T$, and flux, $F$. Consequently, we have $T\propto \dot{M}_{\rm d}/M_{\rm BH}^2$ and $F\propto \dot{M}_{\rm d}/D^2$. This implies that changes in $M_{\rm BH}$, $D$, and $\dot{M}_{\rm d}$ do not affect the fitted spectrum as long as the scaling of $M_{\rm BH} \propto D \propto \dot{M}_{\rm d}^{1/2}$ holds. Thus, to break this degeneracy, we fixed the distance at the best-fit value of \citet{Mata2025A&A...693A.129M}, $D=3.4$ kpc (see also discussion in \citealt{zdziarski2025ApJ...986L..35Z}). This model, \texttt{plabs*tbabs(thcomp*kerrbb+relxillCp)} (now referred to as M5), provides a better fit compared to the previous model M4, with $\chi^2/dof=470.8/520$. The best-fit spectral parameters are given in Table~\ref{tab:Obs2_softstate}, and the corresponding spectral data with the best-fit model are shown in Figure~\ref{fig:softstate_Obs2}.

We obtain the best-fit black hole spin, $a=0.79^{+0.07}_{-0.15}$, mass, $M_{\rm{BH}} = 10.3^{+5.5}_{-2.5}~M_\odot$, and disk inclination angle, $i=46.6^{+6.7}_{-9.2}\degr$. We also estimated the black hole mass using the mass function proposed by \cite{Mata2025A&A...693A.129M}, defined as $f \equiv M_{\rm{BH}} \sin^3i/(1+M_2/M_1)^2 = 2.77\pm0.09M_\odot$, where $M_2$ denotes the mass of the donor star. For this, we adopted the value of inclination angle, $i$, from this model M5 and considered $M_2$ varying between $0.2–0.78~M_\odot$ \citep{Mata2025A&A...693A.129M, zdziarski2025ApJ...986L..35Z}. The black hole mass estimated from the mass function lies in the range $\approx 7$--$12~M_\odot$. Furthermore, the fitted values of $M$ and the disk accretion rate, $\dot{M}_{\rm d}=1.5\pm0.2\times 10^{18}$ g/s, correspond to the disk luminosity of $L_{\rm disk}\approx 0.13 L_{\rm{Edd}}$, assuming the disk accretion efficiency corresponding to the best-fit value of the spin \citep{Cunningham75}. We note here the scalings of $M_{\rm{BH}}\propto D$, $\dot{M}_{\rm d}\propto D^2$ and $L/L_{\rm Edd}\propto D$. The other best-fit spectral parameters, including $\Gamma$, $kT_{\rm e}$, $\log\xi$, and $\log N$, from the model M5 are well consistent with those estimated from the previous model M4.

%%% Xspec spectral plots for soft state data

In our next model, M6, we used the relativistic disk model \texttt{bhspec} to represent the disk's thermal emission, replacing \texttt{kerrbb} in M5. Like \texttt{kerrbb}, the inner disk is assumed to extend down to the ISCO in \texttt{bhspec}. The primary difference between these two relativistic disk models lies in how they treat disk emission. While \texttt{kerrbb} considers a color-corrected blackbody approximation, \texttt{bhspec} employs detailed atmospheric-like calculations for each disk annulus to calculate the vertical disk structure and radiative transfer processes self-consistently \citep{Davis2005ApJ...621..372D, Davis2006ApJS..164..530D}. \texttt{bhspec} is available in two table models, with the viscosity parameters $\alpha=0.01$ and $0.1$ and covering the black hole spin parameters $0-0.99$ and $0-0.8$, respectively. Since $\alpha=0.1$ is more appropriate for BHXRBs, we adopted the second table model even though it allows a smaller range of black hole spin. The black hole spin ($a$) and disk inclination angle ($i$) were tied between \texttt{bhspec} and \texttt{relxillCp}, and allowed to vary. The other free parameters of \texttt{bhspec} are black hole mass ($M_{\rm{BH}}$), and luminosity of the accretion disk ($L/L_{\rm{Edd}}$). We fixed the \texttt{bhspec} normalization, defined as $(10~{\rm{kpc}}/D)^2$, to a value corresponding to $D=3.4$ kpc \citep{Mata2025A&A...693A.129M, zdziarski2025ApJ...986L..35Z}. Allowing the distance to vary resulted in a relatively larger value ($D \gtrsim 5.2$ kpc) and, consequently, a higher black hole mass. This model, \texttt{plabs*tbabs(thcomp*bhspec+relxillCp)}, provided a good and statistically acceptable fit with $\chi^2/dof=489/521$. The best-fit spectral parameters are listed in Table~\ref{tab:Obs2_softstate}, and the spectra, along with the best-fit model, are shown in Figure~\ref{fig:softstate_Obs2}. We find a black hole mass of $M_{\rm{BH}} = 10.2^{+1.3}_{-0.4}~M_\odot$. The obtained values of disk luminosity and inclination angle are consistent with those inferred from M5 within the error bars. Other best-fit spectral parameters are also broadly consistent with those obtained from models M4 and M5.

%%% All PDSs plots in the hard state
\begin{figure}
    \centering
    \includegraphics[width=\columnwidth]{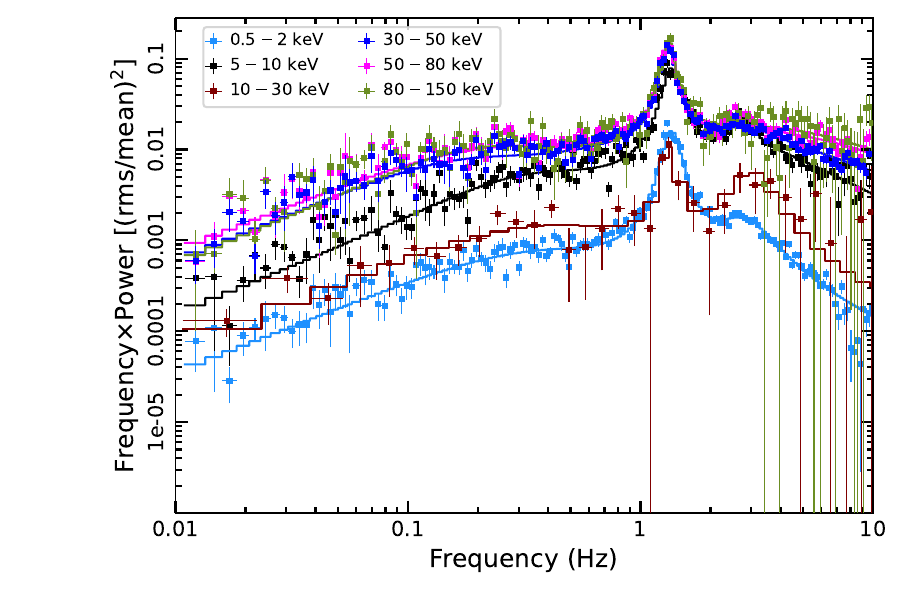}
    \includegraphics[width=\columnwidth]{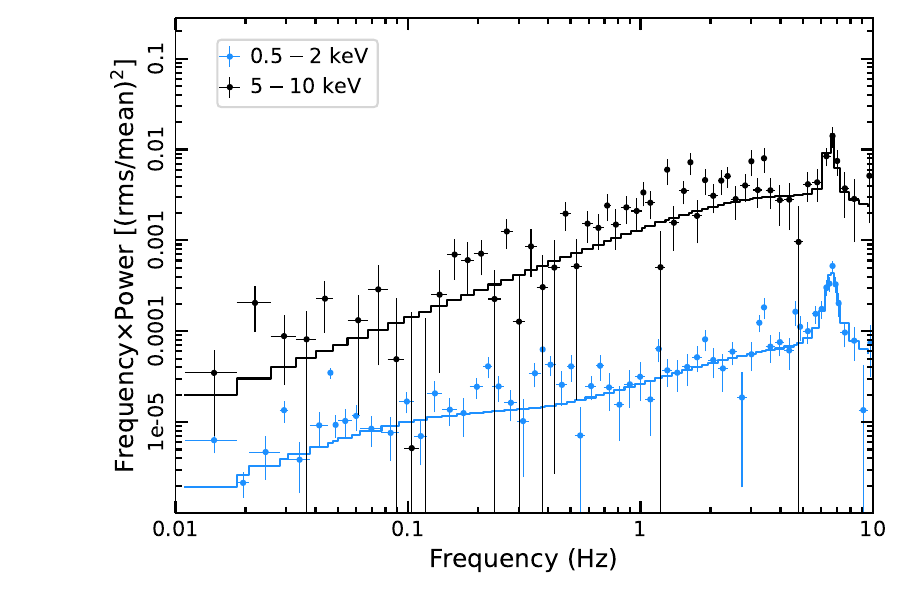}
    \caption{Upper panel: PDSs from the NICER and Insight-HXMT/HE observations, and CPDS from the NuSTAR observation derived in different energy bands during the HIMS. Lower panel: PDSs from the NICER observation derived in two different energy bands during the SIMS. Some of the PDSs are rebinned for plotting purposes only.}
    \label{fig:pds_hardstate_softstate}
\end{figure}

%%% Nicer Fractional RMS
\begin{figure*}
    \centering
    \includegraphics[width=0.45\linewidth]{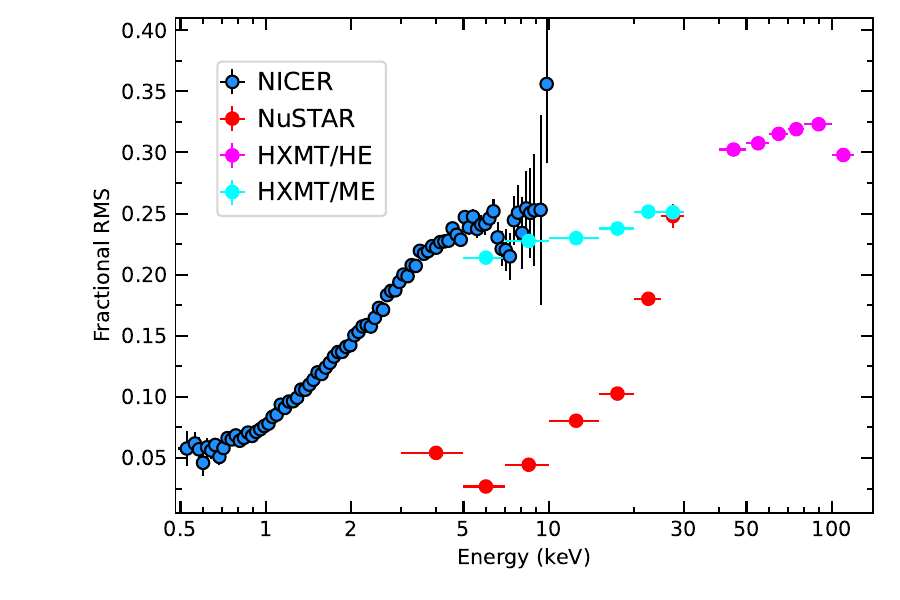}
    \includegraphics[width=0.45\linewidth]{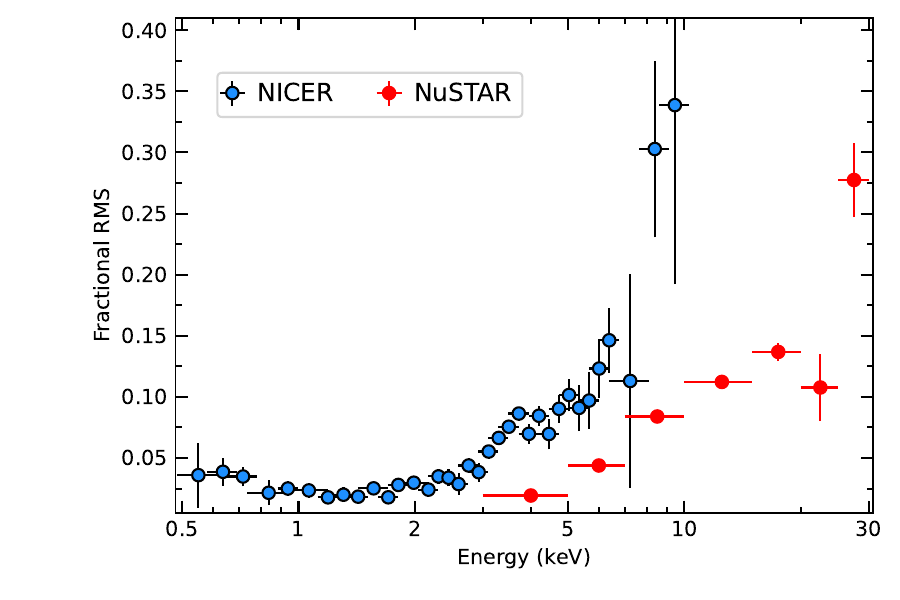}
    \caption{Fractional rms as a function of photon energy for the 0.01-10 Hz range in the HIMS (left panel) and the SIMS (right panel). The blue, red, cyan, and magenta circles represent the NICER, NuSTAR, HXMT/ME, and HXMT/HE observations, respectively.}
    \label{fig:frac_rms}
\end{figure*}

\subsection{Timing Analysis}

To investigate the source variability, we extracted PDSs across multiple energy bands during both the HIMS and SIMS. In the HIMS, we obtained white-noise subtracted and rms-normalized PDSs in the $0.5$–$2$ keV and $5$–-$10$ keV bands from NICER data, and in the $30$–$50$ keV, $50$–$80$ keV, and $80$–-$120$ keV bands from HXMT/HE data using the \texttt{powspec} task of the XRONOS package. For HXMT/HE data, the background-subtracted lightcurves were used to derive the PDSs. Additionally, we derived cross power density spectra (CPDSs) in the $10$–-$30$ keV band from NuSTAR observations using the \texttt{HENDRICS} package \citep{Bachetti2015ApJ...800..109B}, combining the two focal plane modules and subtracting the white noise contribution. The CPDS method involves performing Fourier transforms of two simultaneous light curves from the fpma and fpmb modules and computing their complex cross-spectrum (cospectrum). The real part of this cospectrum serves as a good alternative to a white-noise-subtracted PDS, as it effectively suppresses uncorrelated Poisson noise and correctly handles data gaps \citep[see][for a detailed description]{Bachetti2015ApJ...800..109B}. AstroSat/CZTI data were excluded from the timing analysis due to their low signal-to-noise ratio, likely resulting from the instrument's relatively small effective area and limited sensitivity \citep{Singh2025MNRAS.536.1323S}. In the SIMS, PDSs were derived only from NICER data in the $0.5$–$2$ keV and $5$–$10$ keV bands. The PDSs and CPDS corresponding to both spectral states are presented in Figure~\ref{fig:pds_hardstate_softstate}.

During the HIMS, the power of the PDS at soft energies is lower across all frequencies, increasing gradually with energy and then saturating beyond $\sim30$ keV. Notably, the CPDS in the $10-30$ keV band derived from the NuSTAR observation shows relatively lower power than the PDS in the 5--10 keV band, which could be due to the effect of dead-time in the NuSTAR detectors. Regardless of the energy ranges, a clear presence of a QPO at $\sim1.3$ Hz accompanied by an upper-harmonic at $\sim2.7$ Hz is observed in all the PDSs and CPDS in the HIMS. A similar presence of a QPO, along with an upper harmonic, in the HIMS of this source using AstroSat/LAXPC data was also reported by \citet{Nandi2024MNRAS.531.1149N}. In the SIMS, the overall shape of the PDSs remains similar. However, the QPO frequency has now shifted to a relatively higher frequency of $\sim6.6$ Hz, and any prominent signature of the presence of the upper-harmonic up to $10$ Hz was not noticed. The energy-dependent behavior of the PDSs across both spectral states indicates that low-energy photons, primarily originating from the disk, exhibit significantly lower variability than higher-energy photons, which are dominated by coronal emission.

We also investigated the fractional rms variability amplitude as a function of photon energy in the 0.01--10 Hz frequency range, using both HIMS and SIMS observations. For the HIMS, we derived the rms-energy spectra over the 0.5--120 keV band using data from NICER, NuSTAR, and HXMT/HE, whereas NICER and NuSTAR data were employed in the SIMS to derive the rms-energy spectrum in the $0.5-30$ keV band. We utilized the spectral and timing analysis package \texttt{Stingray} \citep{Huppenkothen2019JOSS....4.1393H, Huppenkothen2019ApJ...881...39H} to derive the rms-energy spectra from NICER data. For the NuSTAR and HXMT/HE data, we first extracted CPDSs and PDSs at different energy bands and then calculated the fractional rms for the $0.01-10$ Hz frequency range from each CPDS and PDS. The rms-energy spectra in two distinct spectral states are shown in Figure~\ref{fig:frac_rms}. In both the HIMS and SIMS, the fractional rms variability amplitude appears to be low ($\sim5\%$) at soft energies, dominated by the thermal disk emission. It gradually increases up to $\sim30-40\%$ at hard energies, where the coronal emission dominates. We note that the NuSTAR data demonstrate significantly lower variability than the NICER data in the same energy bins, likely due to dead time in the NuSTAR detectors, which artificially reduces the variability amplitude \citep[M. Bachetti, private communication;][]{Bachetti2018ApJ...853L..21B, Zdziarski2024ApJ...967L...9Z}. It is worth noting that the effects of dead time are significant for bright sources but become progressively less pronounced for lower-intensity sources. The non-constant and very long ($\sim 2.5$ ms) nature of the NuSTAR dead time further complicates its accurate correction \citep{Bachetti2018ApJ...853L..21B}. To further investigate this effect, we used HXMT/ME data that substantially overlap with the NuSTAR data within a specific energy range. We found that the fractional rms derived from the HXMT/ME data is significantly higher than that from the NuSTAR data, and closely matches the measurements from the NICER data. This further confirms that the low rms in the NuSTAR data is due to dead time.

\section{Discussions} \label{sec:discussion}

We have performed a broadband spectral study of the newly detected Galactic BHXRB Swift~J1727.8--1613 during its 2023 outburst over the $0.7-200$ keV band in the HIMS  using simultaneous NICER, NuSTAR, Insight-HXMT/HE, and AstroSat/CZTI observations, and over the $0.7-78$ keV band in the SIMS using contemporaneous observations from NICER and NuSTAR data. We also investigate source variability between the two spectral states. This outburst is complete, similar to the canonical outbursts observed in low-mass BHXRBs. Substantial evolution in spectral hardness also indicates a transition from the HIMS to the SIMS between the two observations.

\subsection{Thermal Comptonization and Accretion Disk}

The analysis of the NuSTAR/FPMA spectral data reveals that the reflection component was notably weak during both observations (see Figure~\ref{fig:residual_plots}). Although residuals from both observations exhibit a distinct feature in the $6–7$ keV band, indicative of an iron emission line, the characteristic reflection hump, typically peaking in the 20--40 keV band, was either faint or absent. Additionally, an upward trend in the residuals above $\sim50$ keV was observed during the HIMS but was not present during the SIMS. A similar high-energy excess above $40$ keV, along with a weak reflection hump, was previously reported in NuSTAR and Insight-HXMT observations during the LHS by \cite{Liu2024arXiv240603834L}. The authors found the reflection hump in the Insight-HXMT data to be relatively prominent, peaking around 20 keV. They proposed that this feature may be partially artificial, resulting from an improper modeling of the high-energy rollover by the \texttt{cutoffpl} model used in their spectral analysis.

Our spectral analysis in the $0.7–78$ keV band during HIMS indicates that the broadband spectrum requires models that include thermal Comptonization of disk photons in two physically distinct Comptonizing regions, their associated reflections, and thermal emission from the disk. The reflection from the hard component originates from a distant part of the disk, where relativistic effects are weak and strong illumination is observed. On the other hand, the reflection from the soft component, which dominates at low energies, appears more blurred due to its proximity to the inner disk. Polarimetric studies of the source also favor radially extended coronal geometry confined to the disk plane, rather than being aligned along the jet axis \citep{Ingram2024ApJ...968...76I, Veledina2023ApJ...958L..16V}.

 We do not find any evidence for a high-energy tail, typically associated with Comptonization by non-thermal scattering electrons in the corona, up to 78 keV. It is worth mentioning here that \cite{Liu2024arXiv240603834L} attributed the excess above $40$ keV observed in the LHS to the presence of a high-energy tail extending beyond the thermal Comptonization component. Moreover, \cite{Peng2024ApJ...960L..17P} also suggested that the excess above $40$ keV in the NuSTAR and Insight-HXMT spectral data taken in the transitioning period of the source towards the HIMS arises due to the presence of a hard X-ray tail coming from either a relativistic jet or from the base/corona below a slowly moving jet. The authors used a single Comptonizing region and a cutoff power-law model to account for the hard X-ray tail. However, a similar upward trend above $\sim50$ keV (see Figure~\ref{fig:residual_plots}) observed in this work is well reproduced by a model that includes two thermal Comptonizing regions.

Results from the reflection modeling suggest that the inner edge of the accretion disk is truncated well away from the ISCO. Although we can place a lower bound on the inner disk radius ($R_{\rm in}$), this parameter remains relatively insensitive to the specific choice of the continuum model, preventing a definitive constraint on the degree of truncation. The absence of a prominent reflection hump makes it difficult to reliably estimate $R_{\rm in}$ using reflection spectroscopy. However, an independent measurement of $R_{\rm in} \sim 29$--$35~r_{\rm {g}}$ obtained from modeling the disk continuum also supports the scenario of a truncated accretion disk in the HIMS. 

The precise location of the inner edge of the accretion disk in the LHS of BHXRBs remains an open and complex question. Using reflection modeling with \texttt{relxill} and framework of a single Comptonizing region, \cite{Liu2024arXiv240603834L} reported that the accretion disk in Swift~J1727.8--1613 extends close to the ISCO during the LHS. However, their analysis required a super-solar iron abundance, which may be an artifact due to the fixed disk density at $10^{15}~\rm{cm^{-3}}$ in the \texttt{relxill} model \citep{Garcia2018ASPC..515..282G, Jiang2019MNRAS.484.1972J, Zdziarski2022ApJ...928...11Z}. Additionally, they attributed the observed low reflection fraction to an outflowing plasma. Notably, the interpretation of the accretion disk extending close to the ISCO based on the detection of broad Fe K$\alpha$ line in the LHS \citep{2006ApJ...653..525M, 2008ApJ...679L.113M} and the bright LHS \citep{2015ApJ...813...84G, 2019ApJ...885...48G} of a few BHXRBs has been challenged by adopting the framework of double Comptonizing regions \citep{Zdziarski2021ApJ...909L...9Z, Zdziarski2021ApJ...914L...5Z, Zdziarski2022ApJ...928...11Z, Banerjee2024ApJ...964..189B, Chand2024ApJ...972...20C}. In such scenarios, the apparent broadening of the iron line (even at the bright LHS  with $L \approx2-15 \% ~L_{\rm Edd}$) may result from the Compton scattering of the reflected photons within a soft, optically thick Comptonizing region overlaying the disk, rather than the relativistic effect. In addition, \cite{Mahmoud2018MNRAS.480.4040M} showed, using their spectral-timing model including Fourier-resolved spectroscopy, that two distinct Comptonizing regions with a stratified corona are required to describe the shape of the broadband spectra, variability, and energy-dependent lags in the bright LHS of Cyg~X--1. Moreover, \cite{Mahmoud2019MNRAS.486.2137M} demonstrated that the long reverberation lags observed in the LHS of BHXRBs cannot be explained by considering a single homogeneous Comptonizing medium with the accretion disk extending close to the ISCO. Their results suggest that a model with two separate Comptonizing regions and a substantially truncated accretion disk is necessary to account for the long reverberation lags. Apart from this, \citet{GarciaF2021MNRAS.501.3173G} argued that the radiative properties of type-B QPOs in the 0.8--10 keV band during the SIMS of MAXI~J1348–630 are best explained by two physically connected Comptonizing regions.

The moderate-to-high disk ionization inferred from our analysis is likely due to intense coronal illumination. However, the relatively high intrinsic disk temperature ($\sim0.4$ keV) may also contribute significantly to the elevated ionization state. The covering fraction ($C_{\rm{frac}} \sim 0.4$) suggests that a substantial portion of the disk photons is intercepted by the soft Comptonizing region, further supporting its interpretation as a warm, optically thick layer enveloping the disk. The overall spectral characteristics observed in the HIMS of Swift~J1727.8--1613 closely resemble the broadband ($0.7–100$ keV) spectrum of the low-mass BHXRB GX~339--4 in the same state \citep{Chand2024ApJ...972...20C}.

\subsection{Hybrid Comptonization}

We observe that Comptonization by a purely thermal electron population fails to account for the high-energy excess at $\gtrsim100$ keV (see Figure~\ref{fig:res_he_czti}); instead, a hybrid electron distribution is required. In particular, $\gamma_{\rm {min}} \sim 1.3$, with electron power-law index of $2$, suggests that a significant portion of the high-energy emission arises from inverse Compton scattering of soft photons by the non-thermally distributed electrons. This provides strong evidence for a hybrid coronal plasma, in which thermal electrons dominate the Comptonized spectrum at low energies, with a non-thermal tail emerging above a specific energy due to non-thermal electron populations. This hybrid model also favors the scenario in which the disk is truncated far from the ISCO, consistent with non-thermal particle acceleration in a hot, optically thin inner accretion flow.

Evidence of the possible presence of a hard X-ray tail originating from non-thermal electrons in the LHS of Swift~J1727.8--1613 during the same outburst was previously reported by \cite{Liu2024arXiv240603834L}. Their results show that the contribution from the non-thermal electron population was relatively soft, with a steeper distribution at the initial phase of the outburst. That particle acceleration gradually increased as the source transitioned towards the HIMS during the rising phase. This is consistent with our current results, which show significant non-thermal particle acceleration in the HIMS. Furthermore, hybrid Comptonization with a significant non-thermal electron population with $\gamma_{\rm{min}}~\sim1.3$ and slightly steep electron power-law index of $\sim2.5$ was also reported in the LHS of the bright low-mass BHXRB MAXI~J1820+070 by \cite{Zdziarski2021ApJ...914L...5Z}. This detection, along with a substantially truncated accretion disk in the LHS, was further confirmed in a follow-up study that included Insight-HXMT data by \citet{Zdziarski2022ApJ...928...11Z}.

 The electron temperature, $kT_{\rm{e}} \sim13-18$ keV, of the hard Comptonizing region, inferred from all three spectral models (M1-M3) in the HIMS, remains low, which could be a consequence of the energy balance within the accretion flow. Alternatively, such a low $kT_{\rm e}$ could also result from efficient cooling by synchrotron photons produced by a hybrid electron distribution \citep{Zdziarski2004PThPS.155...99Z, Malzac2009MNRAS.392..570M, Veledina2011ApJ...737L..17V, Zdziarski2021ApJ...914L...5Z}.

\subsection{State Transition and Disk Evolution}

In contrast to the HIMS, the coronal structure during Obs2 appears relatively simple and can be described by a single Comptonizing region. A relatively steeper photon index and a low covering fraction further indicate that the corona became weak in this state. We also observe a significant evolution in the accretion disk, with elevated inner-disk temperature and ionization, and a relatively closer extent to the ISCO (as inferred from reflection and disk-continuum modeling). These further suggest that the source transitioned to the SIMS during Obs2.

\subsection{Spin, Mass, Distance, and Disk Inclination}

Reflection modeling alone (see model M4) could not constrain the spin, yielding a value $\gtrsim 0.7$ at the $90\%$ confidence level. \citet{Peng2024ApJ...960L..17P} reported a very high black hole spin of $0.98^{+0.02}_{-0.07}$ for Swift~J1727.8--1613 by modeling the reflection continuum in the LHS. However, the reflection method for spin measurement relies on the assumption that the disk extends down to the ISCO or very close to it, and the extent of the inner accretion disk in the LHS is highly debated. On the other hand, our spin estimation of $0.79^{+0.07}_{-0.15}$ by modeling the disk continuum and reflection component together (see model M5) is slightly lower but broadly consistent with the value reported by \citet{Svoboda2024ApJ...966L..35S}, using the HSS observations of the source in February 2024.

We find a close agreement between our estimated black hole mass, $M_{\rm{BH}} \approx 10~M_\odot$ (see model M5 and M6), and previously reported values. This value also appears to be consistent with the black hole mass obtained from the mass function suggested by \cite{Mata2025A&A...693A.129M}. Furthermore, \citet{zdziarski2025ApJ...986L..35Z} favored a black hole mass of $\lesssim 13M_\odot$ for Swift~J1727.8--1613, which aligns well with our estimate. Additionally, \citet{Debnath2025arXiv250416391D}, using the propagating oscillatory shock model and the observed QPO–photon index correlation, reported a black hole mass of $\sim13~M_\odot$ for this source that is also consistent with our results. It should be noted here that the \texttt{bhspec} model does not account for returning radiation, in which photons emitted from the inner disk are gravitationally bent back onto the disk surface. This process can introduce additional heating and alter the observed thermal spectrum, particularly in the high-energy tail. The effect becomes significant for systems with high black hole spin and high inclination. Thus, omission of the effect of returning radiation in the \texttt{bhspec} model may introduce a systematic uncertainty in the derived disk parameters. 

The disk inclination angle obtained from the models M5 and M6 lies in the range of $\sim 37\degr$–-$53\degr$, which is in good agreement with the $30\degr$--$50\degr$  range estimated by \citet{Svoboda2024ApJ...966L..35S} from polarimetric analysis of the HSS observation. Additionally, \citet{Peng2024ApJ...960L..17P} also reported a disk inclination angle of $\sim40\degr$  using LHS observations in 2023. However, some studies reported the disk inclination angle in Swift~J1727.8--1613 to be $\gtrsim78\degr$  \citep{Chatterjee2024ApJ...977..148C, Debnath2024ApJ...975..194D}, which is unlikely given the absence of any eclipse or dips in the source lightcurve.

\subsection{Source Variability}

The shapes of the PDSs in different energy bands indicate a slight decrease in variability as the source transitioned to the SIMS. In both spectral states, the increase in PDS power with energy suggests that high-energy coronal photons exhibit greater variability than the softer ones associated with the disk. In the HIMS, however, the PDSs above $30$ keV are nearly identical, suggesting a similar variability pattern among hard photons in this energy range. Additionally, the significant increase in QPO frequency from $\sim1.3$ Hz to $\sim6.6$ Hz suggests a change in the disk–corona geometry between the two states. If QPOs are thought to be associated with the inner edge of the disk or the outer boundary of the inner hot flow, the observed increase in the QPO frequency implies that the disk has moved relatively closer to the ISCO in the SIMS, potentially accompanied by a contraction of the corona. This interpretation is consistent with the results of our spectral modeling, which show that the disk moves closer to the ISCO in the SIMS.

 The shapes of the rms–energy spectra observed in the HIMS and SIMS differ from those reported by \citet{Gierlinski2005MNRAS.363.1349G} for the LHS of XTE~J1650--500 using RXTE data, where the variability is more substantial at lower energies and gradually decreases with increasing photon energy. However, their study lacked coverage below $3$ keV, making it difficult to investigate the variability associated with the disk emission at softer energies. The morphology of the rms-energy spectra does not always remain constant and varies across spectral states and sources. In the intermediate states of the BHXRB XTE~J1550--564, \cite{Gierlinski2005MNRAS.363.1349G} reported that fractional rms increases with the increase in photon energy. This trend is consistent with our findings from both HIMS and SIMS for Swift~J1727.8--1613. In addition, a similar trend in the rms–energy spectrum derived at the QPO frequency was also reported during the HIMS of the low-mass BHXRB MAXI~J1803--298 by \cite{Chand2022ApJ...933...69C}, though their study also lacked the data below $3$ keV.

 Our investigation of the rms-energy spectra in both spectral states reveals a weakly variable or stable disk and a highly variable power-law component, which primarily governs the overall source variability. The nature of the variability is similar across the two states, although a slight decrease is observed in the SIMS. Since our HIMS data are limited to $120$ keV, and our spectral modeling indicates that the contribution from the non-thermal distribution becomes significant above $\sim100$ keV, a robust statement about the variability associated with the non-thermal electron population is difficult. Moreover, the lack of high-energy data in the SIMS also prevents us from studying the variability pattern driven by high-energy photons ($>$30 keV) and the non-thermal electron population in this state.

\section{Conclusions} \label{sec:conclusion}

Our main results from this study are as follows.

The broadband spectrum up to $78$ keV in HIMS can be described by models with two Comptonizing regions, in which the electron distribution is predominantly thermal (Maxwellian). We did not observe any contribution from the non-thermal electron population in the HIMS spectral data up to 78 keV. The disk-corona geometry of Swift~J1727.8--1613 shows substantial similarity with that of GX~339--4 in the HIMS \citep{Chand2024ApJ...972...20C}.

The contributions from the non-thermal electron distribution become significant above 100 keV, where a weak hard X-ray tail is observed. The shape of this tail can be satisfactorily described considering a hybrid electron distribution in the corona, where the electrons are thermally populated up to a specific energy, and then become non-thermal beyond the thermal cutoff.

The low electron temperature in the hard Comptonizing region could result from an energy balance within the hot flow or from efficient cooling by synchrotron photons produced by a hybrid electron plasma, which can prevent electrons from reaching higher temperatures.

We find from reflection spectroscopy that the accretion disk is likely truncated by at least $33 r_{\rm{g}}$ from the ISCO in the HIMS, which also agrees closely with values estimated from modeling the disk continuum. The truncated disk scenario is consistent with previous results for the bright LHS of other low-mass BHXRBs within the framework of dual Comptonizing regions \citep{Zdziarski2021ApJ...909L...9Z, Banerjee2024ApJ...964..189B, Chand2024ApJ...972...20C}.

The source transitions from HIMS in Obs1 to SIMS in Obs2, with an elevated disk temperature and a relatively steeper photon index. Unlike in HIMS, the requirement for a single Comptonizing region to describe the broadband spectrum in SIMS suggests a significant change in coronal geometry across the state transition. This interpretation is further supported by the disk's approach to the ISCO and by a substantial increase in the QPO frequency during the SIMS. However, the lack of high-energy data prevents us from probing the existence of a hard X-ray tail that may originate from the acceleration of a non-thermal electron population.

Our estimates of the black hole mass, spin, and disk inclination angle, obtained by jointly modeling the disk continuum and reflection components, broadly agree with previously reported values. Despite the uncertainties, these parameters are consistent across the models. 

The shape of the rms–energy spectra is consistent with those observed in the intermediate states of other low-mass BHXRBs. Our results indicate that the accretion disk remains relatively stable or only moderately variable, while the Comptonizing component exhibits substantial variability in both spectral states. This behavior suggests that the Comptonizing region primarily governs the overall source variability.

\section{Acknowledgments}
We thank the anonymous reviewer for their comments, which have helped improve the quality of this paper. This research used archival data from the NICER and NuSTAR missions, obtained from the High Energy Astrophysics Science Archive Research Center (HEASARC) and provided by NASA's Goddard Space Flight Center. It also utilizes data from the CZTI instrument onboard AstroSat, a mission of the Indian Space Research Organisation (ISRO), archived at the Indian Space Science Data Centre (ISSDC). This work used data and software from the Insight-HXMT mission, a project funded by the China National Space Administration (CNSA) and the Chinese Academy of Sciences (CAS). We also acknowledge the use of MAXI mission data, provided by RIKEN, JAXA, and the MAXI team. We thank N. P. S. Mithun from the CZTI team for valuable discussions on the AstroSat/CZTI background. SC thanks Hsiang-Kuang Chang for his support. SC acknowledges support from the Polish Academy of Sciences through their study visit program. AAZ acknowledges support from the Polish National Science Center grants 2019/35/B/ST9/03944 and 2023/48/Q/ST9/00138.

\vspace{5mm}
\facilities{NICER, NuSTAR, AstroSat, Insight-HXMT, and MAXI.}

\software{HEASoft \citep[v.6.34;][]{heasoft2014ascl.soft08004N}, XSPEC \citep[v.12.14.1;][]{Arnaud1996ASPC..101...17A}, CZTPIPELINE (v.3.0.1; \url{http://astrosat-ssc.iucaa.in/cztiData}), HXMTDAS \citep[v.2.06;][]{Zhao2020ASPC..527..469Z}, Stingray \citep{Huppenkothen2019JOSS....4.1393H, Huppenkothen2019ApJ...881...39H}, Matplotlib \citep{Hunter2007CSE.....9...90H}.}

\bibliography{reference}{}
\bibliographystyle{aasjournal}

\end{document}